\documentclass[preprint,review,12pt,3p]{elsarticle}
\usepackage{graphicx,subfigure}
\usepackage{comment}
\usepackage{ctable}
\usepackage{caption}
\usepackage{color}

\usepackage{amssymb}
\usepackage{amsmath}
\usepackage{bbm}
\usepackage{wrapfig}
\usepackage{cases}

\usepackage[english]{babel}
\usepackage{amsmath,amsthm}
\usepackage{amsfonts}
 \usepackage{amsthm}
 \usepackage{float}
 \usepackage{epstopdf}

\theoremstyle{definition}

\theoremstyle{remark}

\numberwithin{equation}{section}

\newcommand{\D}{\mathrm{d}}

\journal{}

\begin{document}

\title{Dislocation dynamics formulation for self-climb of dislocation loops by vacancy pipe diffusion}%

 \author[xiamen,hkust]{Xiaohua Niu}
\author[nano]{Yejun Gu}
 \author[hkust]{Yang Xiang\corref{cor1}}
\ead{maxiang@ust.hk}

\address[xiamen]{School of Applied Mathematics,  Xiamen University of Technology,  Xiamen,   361024,   China. }

\address[hkust]{Department of Mathematics, The Hong Kong University of Science and Technology, Clear Water Bay, Kowloon, Hong Kong}

\cortext[cor1]{Corresponding author}

\address[nano]{Department of Mechanical Engineering, Whiting School of Engineering, The Johns Hopkins University, Baltimore, MD 21218, USA}

\begin{abstract}
It has been shown in experiments that self-climb of prismatic dislocation loops by pipe diffusion plays important roles in their dynamical behaviors, e.g., coarsening of prismatic loops upon annealing, as well as the physical and mechanical properties of materials with irradiation.
In this paper, we show that this dislocation dynamics self-climb formulation that we derived in Ref.~\cite{Niu} is able to quantitatively describe the properties of self-climb of prismatic loops that were observed in experiments and atomistic simulations. This dislocation dynamics formulation applies to self-climb by pipe diffusion for any configurations of dislocations, and is able to recover the available models in the literature for rigid self-climb motion of small prismatic loops.
We also present DDD implementation method of this self-climb formulation. Simulations performed show evolution, translation  and coalescence of prismatic loops
  as well as prismatic loops driven by an edge dislocation by  self-climb motion and the elastic interaction between them.
 These results are in excellent agreement with available experimental and atomistic results.
We have also performed systematic analyses of the behaviors of a prismatic loop under the elastic interaction with an infinite, straight edge dislocation by motions of self-climb and glide.

\end{abstract}


\maketitle
\section{Introduction}

Prismatic dislocation loops are often formed by the precipitation and collapse  of excess point-defects of vacancies or interstitials that are produced by quenching or by fast-particle irradiation \cite{Hirth-Lothe,Burton1986,Was,DudarevReview}.  The Burgers vector of a prismatic dislocation loop has a component normal to its habit plane. Formation and evolution of these prismatic loops crucially influences the physical and mechanical properties of the materials.

It was observed in the early experiments  \cite{Kroupa1961}  that  circular prismatic dislocation loops are able to climb out their glide cylinders driven by repulsive interaction with edge dislocations with different Burgers vectors. The enclosed area of the such a climbing prismatic loop was found to be conserved as the loop moves, thus  it cannot be attributed to dislocation climb assisted by vacancy bulk diffusion which makes the prismatic loops shrink. This climb motion of prismatic loops was called {\it conservative climb} of dislocation loops and was explained in terms of pipe diffusion along the loops \cite{Kroupa1961,Hirth-Lothe}. It has also been observed in early experiments that the number of prismatic dislocation loops in quenched materials decreases while the size of the loops increases upon annealing \cite{Silcox1960,Washburn1960}. These experimental results were explained later on based on the mechanism of motion and coalescence of dislocation loops by pipe diffusion, which is called the {\it self-climb} mechanism \cite{Johnson1960}. More details of the coalescence have been observed in the later experiments \cite{Turnbull1970,Narayan1972,Swinburne2016}.  It was demonstrated in Ref.~\cite{Swinburne2016} by comparisons of predictions of their atomistic self-climb model with TEM observations that self-climb by pipe diffusion is the dominant mechanism of prismatic loop coalescence at not very high temperatures. These authors showed
that two nearby prismatic loops of order $10$nm in iron at 750K coalesce after a time of order $10$s in both results of their atomistic self-climb model and experiment, and it is six orders of magnitude faster than the time predicted by using bulk vacancy diffusion assisted climb. Such self-climb (conservative climb) motion of dislocations  plays critical roles in the properties of irradiated materials and has been an active research topic ever since it was first observed in 1960's \cite{Hirth-Lothe,Kroupa1961,Silcox1960,Washburn1960,Johnson1960,Turnbull1970,Narayan1972,Burton1986,DudarevReview,
Swinburne2016,DudarevLangevin,Okita2016,Hayakawa2016}.

Available quantitative theories in the literature of the self-climb of dislocation loops are all based on the dynamics of small circular loops whose shape do not change during the evolution, and the dynamics of each circular loop is driven by the interaction for between the loop and an external stress gradient following a linearly mobility relation~\cite{Johnson1960,Kroupa1961,Turnbull1970,Narayan1972,Swinburne2016,Okita2016}. Parameters in these theories were obtained by comparisons with experimental results \cite{Johnson1960,Turnbull1970,Narayan1972,Swinburne2016} or atomistic simulations \cite{Swinburne2016,Okita2016}. Using models within this loop dynamics framework, self-climb assisted coalescence and coarsening of prismatic loops were simulated and the results of the former were compared with those of experiments \cite{Swinburne2016}; behavior of a self-interstitial dislocation loop near an edge dislocation in BCC-Fe was studied \cite{Okita2016,Hayakawa2016}; spatial ordering of nano-dislocation loops in ion-irradiated materials was investigated by Langevin dynamics simulations of interacting dislocation loops \cite{DudarevLangevin}. A recent molecular dynamics simulations of nanoindentation showed that half prismatic edge dislocation loops annihilate via pipe diffusion of vacancies from the free surface  \cite{Mordehai2017}.
Despite these research progresses on the understanding of dislocation self-climb mechanisms and related properties, a simulation model based on discrete dislocation dynamics (DDD) that gives the self-climb velocity at each point of the dislocations, as those available DDD simulation methods for vacancy bulk diffusion assisted climb, e.g. \cite{Ghoniem,Xiang2003,Xiang2006,Arsenlis,Mordehai,Bako,Keralavarma,Ayas2014,HuangMS2014,Gu2015}, is still lacking.  Although a pipe-diffusion-based dislocation climb DDD model was proposed in Ref.~\cite{GaoIJP2011} in which the climb velocity is proportional to the vacancy flux along the dislocation, their simulations showed that prismatic loops shrink, thus their model is not able to describe the conservative climb of prismatic loops observed in the experiments and atomistic simulations reviewed above, in which the area enclosed by a prismatic loop is conserved during the self-climb motion.

      Recently, we have derived a dislocation climb model within the DDD framework \cite{Niu} from  a stochastic scheme on the atomistic scale, where both the vacancy bulk diffusion assisted climb and the self-climb assisted by pipe diffusion are considered and the total climb velocity is given at any point on the dislocations. Preliminary examination showed that the derived self-climb velocity formula is able to maintain the area enclosed by a prismatic loop as it moves by self-climb. It has also been shown that this self-climb formulation is able to quantitatively explain the half prismatic edge dislocation loops annihilation with free surface via pipe diffusion of vacancies observed in MD simulations \cite{Mordehai2017} and self-healing of low angle grain boundaries by vacancy diffusion \cite{Gu2018}.

      In this paper, we show that our self-climb formulation derived in Ref.~\cite{Niu} is able to quantitatively describe the properties of self-climb of prismatic loops that were observed in experiments and atomistic simulations. For small circular prismatic loops, our formulation is able to recover the available models in the literature based on linearly mobility relation driven by the interaction force between the loops and an external stress gradient.
      We also present DDD implementation method of this self-climb formulation.
      DDD Simulations are performed  for the motion and interaction of prismatic loops and interaction between an edge dislocation and prismatic loops. Especially, our simulations show
       prismatic loops driven by an edge dislocation and coalescence of prismatic loops by the self-climb motion, which is in excellent agreement with the experimental observations \cite{Kroupa1961} (prismatic loops driven by an edge dislocation) and \cite{Silcox1960,Washburn1960,Johnson1960} (coalescence of prismatic loops). We have also performed systematic analyses of the behaviors of a prismatic loop under the elastic interaction with an infinite, straight edge dislocation by motions of self-climb and glide (including both rotation and rigid motions by glide) of the prismatic loop.

This paper is organized as follows. In Sec.~\ref{sec:formulation}, we review our self-climb formulation derived in Ref.~\cite{Niu} and discuss its physical implication. In Sec.~\ref{sec:DDD}, we present a numerical implementation method of the self-climb formulation. In Sec.~\ref{sec:properties}, we study the properties of self-climb  of prismatic loops based on this formulation, including properties of conservation of enclosed area, equilibrium shape, and translation velocity with stress gradient, etc.
In Sec.~\ref{sec:simulations},  we perform  DDD simulations of the self-climb of dislocations  and compare the results with experimental observations and atomistic simulations. In Sec.~\ref{sec:interaction}, we perform systematic analyses of the behaviors of a prismatic loop under the elastic interaction with an infinite, straight edge dislocation by motions of self-climb and glide.

\section{Dislocation self-climb in DDD simulations}\label{sec:formulation}

In this section, we briefly review the dislocation climb formulation that we derived in Ref.~\cite{Niu} from atomistic schemes and discuss its physical implication. In this formulation, the dislocation climb velocity consists of contributions from both the climb due to vacancy diffusion in the bulk and the climb due to vacancy pipe diffusion along the dislocations, i.e.
\begin{equation}
v_{\rm cl}=v_{\rm cl}^{\rm (b)}+v_{\rm cl}^{\rm (p)},
\end{equation}
where $v_{\rm cl}^{\rm (b)}$ and $v_{\rm cl}^{\rm (p)}$ are the dislocation climb velocities due to vacancy bulk diffusion and pipe diffusion, respectively. Dislocation climb due to vacancy pipe diffusion is also referred to as the self-climb in the literature, and accordingly, $v_{\rm cl}^{\rm (p)}$ is also called the self-climb velocity.
The DDD study on dislocation climb in the literature was mainly focused on the climb due to vacancy bulk diffusion as briefly reviewed in the introduction, whose contribution to the total dislocation climb velocity is  $v_{\rm cl}^{\rm (b)}$. In this paper, we focus on the DDD implementation of the dislocation self-climb velocity $v_{\rm cl}^{\rm (p)}$ and the related properties.

The DDD self-climb velocity derived in Ref.~\cite{Niu} is
\begin{equation} \label{eq:climb-v}
v_{\rm cl}^{\rm (p)}=c_0D_cb\frac{\D^2 }{\D s^2}\left(e^{-\frac{f_{\rm cl}\Omega}{bk_BT}}\right),
\end{equation}
where  $s$ is the arclength parameter along the dislocation, $b$ is the length of the Burgers vector of the dislocation, $c_0$ is the reference equilibrium vacancy concentration on the dislocation without climb force, $D_c$ is the vacancy pipe diffusion constant, $k_B$ is Boltzmann constant, $T$ is the temperature, $\Omega$ is the atomic volume, and $f_{\rm cl}$ is the climb component of the Peach-Koehler force on the dislocation, which is  given by
\begin{equation}\label{eqn:fcl0}
f_{\rm cl}=\mathbf{f}_{\rm PK}\cdot \mathbf l_{\rm cl}.
\end{equation}
Here
$\mathbf{f}_{\rm PK}$ is the Peach-Koehler force calculated by
$\mathbf{f}_{\rm PK}=(\boldsymbol{\sigma}\cdot \mathbf{b})\times \boldsymbol{\xi}$~\cite{Hirth-Lothe},  $\mathbf b$ is the Burgers vector of the dislocation, $\boldsymbol{\sigma}$ is the stress tensor, $\boldsymbol{\xi}$ is the dislocation line direction, and
$\mathbf l_{\rm cl}$ is the climb direction  defined as $\mathbf l_{\rm cl}=\boldsymbol{\xi}\times\mathbf{b}/b$.
Accordingly, we have
\begin{equation}\label{climbdirections}
\mathbf v_{\rm cl}^{\rm (p)}=v_{\rm cl}^{\rm (p)}\mathbf l_{\rm cl}.
\end{equation}
 For a prismatic loop, $\boldsymbol{\xi}$ is always normal to $\mathbf b$, the climb force can be simplified as:
\begin{equation}\label{eqn:fcl}
f_{\rm cl}=-\frac{1}{b}(\mathbf{b}\cdot\boldsymbol{\sigma}\cdot \mathbf{b}).
\end{equation}

As in the climb assisted by vacancy bulk diffusion (e.g. \cite{Hirth-Lothe,Mordehai,Bako,Keralavarma,Ayas2014,Gu2015}), when $f_{\rm cl}<<bk_BT/\Omega$, $e^{-\frac{f_{\rm cl}\Omega}{bk_BT}}\approx 1-\frac{f_{\rm cl}\Omega}{bk_BT}$, and the self-climb velocity  in Eq.~\eqref{eq:climb-v} can be simplified as
\begin{equation} \label{eq:climb-v0}
v_{\rm cl}^{\rm (p)}=-\frac{c_0D_c\Omega}{k_BT}\frac{\D^2 f_{\rm cl}}{\D s^2}.
\end{equation}

\noindent
{\bf Remarks:}
\begin{itemize}
\item
In Eq.~\eqref{eq:climb-v},  the reference equilibrium vacancy concentration on the dislocation $c_0$ and  the vacancy pipe diffusion constant $D_c$ may also depend on the dislocation character.  In this case, Eq.~\eqref{eq:climb-v} can be written as
$v_{\rm cl}^{\rm (p)}=D_cb\frac{\D^2 }{\D s^2}\left(c_0e^{-\frac{f_{\rm cl}\Omega}{bk_BT}}\right)$. As in almost all the available studies for dislocation climb, we use the definition of the vacancy standard state $c_0$  in Ref.~\cite{Hirth-Lothe}. A different definition of the vacancy standard state $c_0$ was adopted by Weertman \cite{Weertman1965,Landau}, leading to an additional contribution to the climb force. The two definitions lead to equivalent climb formulations, see the discussion in Ref.~\cite{Hirth-Lothe}.
\item
 In some studies in the literature, line tension approximation of the force on a dislocation loop was adopted for simplification in calculations. Under the line tension approximation and without applied stress, Eq.~\eqref{eq:climb-v0} becomes
$v_{\rm cl}^{\rm (p)}=-\frac{c_0D_c\Omega}{k_BT}\frac{\D^2 }{\D s^2}(T_n\kappa)$, where $\kappa$ is the curvature of the loop and $T_n$ is the line tension. For a prismatic loop, the line tension is $T_n=\frac{\mu b^2}{4\pi(1-\nu)}\ln\frac{L_c}{r_d}$, where $r_d$ is the dislocation core radius, $L_c$ is the outer cutoff length and $L_c=8R/e$ for a circular loop with radius $R$ \cite{Hirth-Lothe}. This approximation is not used in this paper.
\end{itemize}

 This dislocation dynamics model for vacancy-assisted dislocation climb is derived by upscaling in space and time from the jog dynamics model. (See Ref.~\cite{Niu} for detailed derivation.) It incorporates the jogged structure and vacancy pipe diffusion and adsorption/emission at jogs, and does not need to solve the vacancy pipe diffusion on the dislocation or track the location and motion of individual jogs.  In the derivation, we have assumed that:
\begin{enumerate}[(i)]
      \item
 The jogs on dislocations are sparsely distributed on the atomic length scale, and the average distance between two adjacent jogs is much smaller than the size of the dislocation loop in the dislocation dynamics model.

 \item
 The vacancy pipe diffusion is much faster than the vacancy bulk diffusion and the vacancy exchange between the dislocation core and the bulk.

 \item
 A jog moves forward or backward when it absorbs or emits a vacancy.

  \item
  The equilibrium vacancy concentration at a jog site is  $c_0e^{-\frac{f_{\rm cl}\Omega}{bkT}}$, where $c_0$ is the reference equilibrium vacancy concentration on the dislocation.



  \end{enumerate}

\begin{figure}[!htbp]
\centering
 \includegraphics [width=0.8\textwidth]{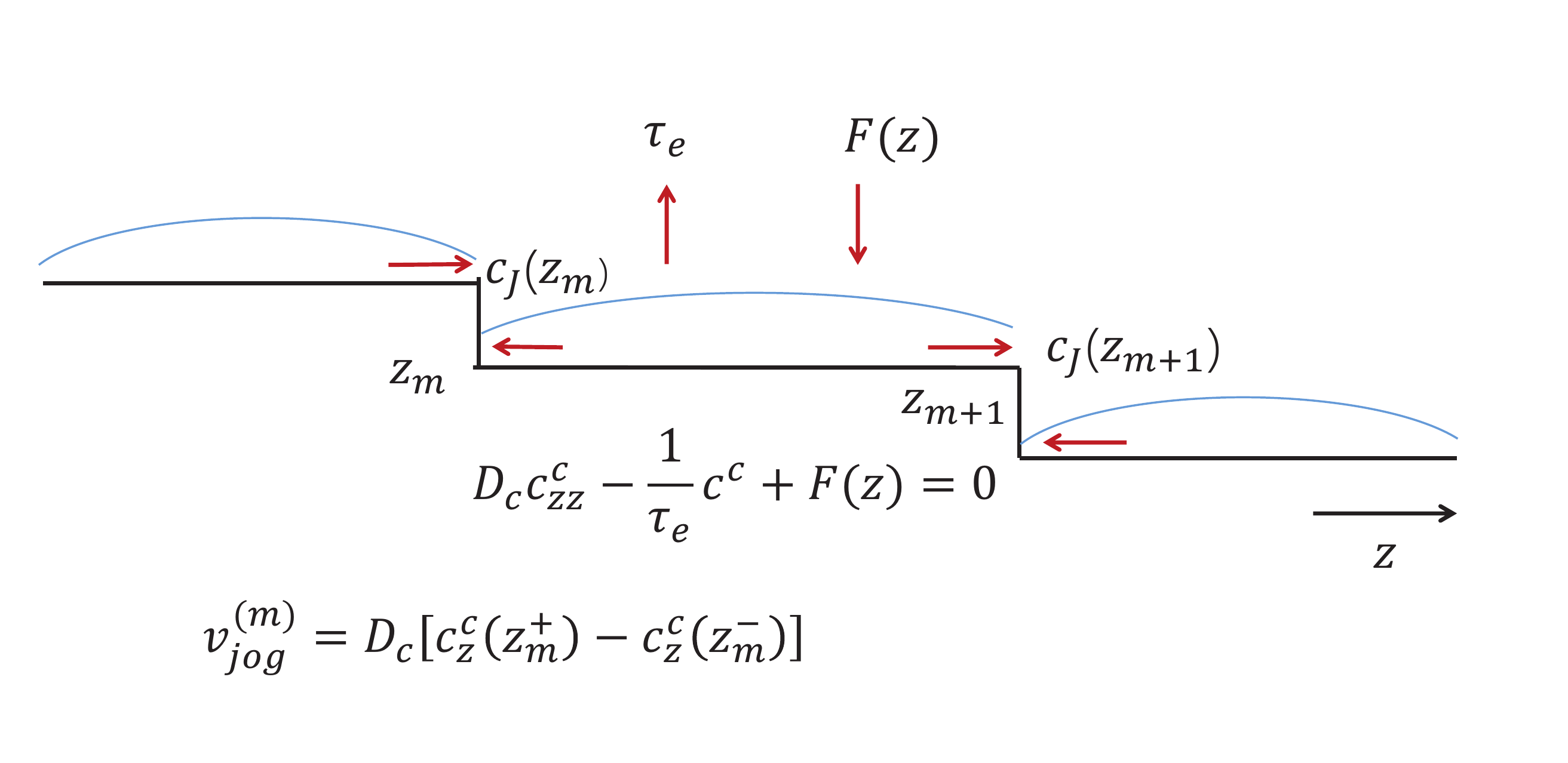}
\caption{The underlying jog dynamics model from which the self-climb DDD model in Eq.~\eqref{eq:climb-v}  is derived. In the DDD model, the upscaled dislocation profile is a smooth curve without explicit jog structure, as shown by the smooth blue curve. }
\label{fig:jog}
\end{figure}

Here we briefly review the derivation of the self-climb DDD model in Eq.~\eqref{eq:climb-v} to show more of its underlying physics.  The underlying jog dynamics model when only vacancy pipe diffusion is considered, from which the self-climb DDD model in Eq.~\eqref{eq:climb-v} is derived, is
\begin{eqnarray}
    \left\{
        \begin{array}{l}
            c^c_t=D_cc^c_{zz}, \ z\in (z_m,z_{m+1}),\vspace{1ex} \\
            c^c(z_m)=\left.c_0e^{-\frac{f_{\rm cl}\Omega}{bk_BT}}\right|_{z=z_m}, \ \  c^c(z_{m+1})=\left.c_0e^{-\frac{f_{\rm cl}\Omega}{bk_BT}}\right|_{z=z_{m+1}},
        \end{array}
    \right.
\end{eqnarray}
where $c^c$ is the vacancy concentration on the dislocation, $z$ is the coordinate along the dislocation, $c^c_t=\frac{\partial c^c}{\partial t}$, $c^c_z=\frac{\partial c^c}{\partial z}$, $c^c_{zz}=\frac{\partial^2 c^c}{\partial z^2}$ are partial derivatives, $D_c$ is the vacancy pipe diffusion constant,
 $z=z_m$, $m$ integers, are locations of the jog on the dislocation,
  $c_0e^{-\frac{f_{\rm cl}\Omega}{bk_BT}}|_{z=z_m}$ is the equilibrium vacancy concentration at the jog $z_m$,  and  $c_0$ is the equilibrium vacancy concentration on the dislocation without climb force.
The velocity of the jog located at $z=z_m$ is
\begin{eqnarray}\label{eqn:vjog}
  v_{\rm jog}^{(m)}
  =D_c\big[c^c_z(z_m^+)-c^c_z(z_m^-)\big].
  \end{eqnarray}
See Fig.~\ref{fig:jog} for an illustration.
 Under the assumptions, we have  $c^c_t=D_cc^c_{zz}\approx 0$ on a dislocation segment between two adjacent jogs.
 The vacancy concentration $c^c$ on each dislocation segment can be solved from
this diffusion equilibrium equation on it, which is $c^c(z)=A_m(z-z_m)+c^c(z_m)$, in $[z_{m},z_{m+1}]$, where the coefficient $A_m=(c^c(z_{m+1})-c^c(z_m))/l_m$ and the length of this dislocation segment $l_m=z_{m+1}-z_m$. It can be calculated using this solution and Eq.~\eqref{eqn:vjog} that the jog velocity $v_{\rm jog}^{(m)}=D_c(A_{m}-A_{m-1})$.  In the DDD model, a dislocation is a smooth curve without explicit jog structure (see Fig.~\ref{fig:jog}), after upscaling from the jog dynamics model, the self-climb velocity is
\begin{eqnarray}\label{eqn:vjogtoclimb}
 v_{\rm cl}(z_m)=\frac{b}{(l_m+l_{m+1})/2}v_{\rm jog}^{(m)},
  \end{eqnarray}
  which leads to Eq.~\eqref{eq:climb-v}.
See Ref.~\cite{Niu} for detailed derivation, which also accounts for dislocation climb assisted by vacancy bulk diffusion.

\section{Numerical implementation of the self-climb formulation}\label{sec:DDD}

\begin{figure}[!htbp]
\centering
 \includegraphics [width=8cm]{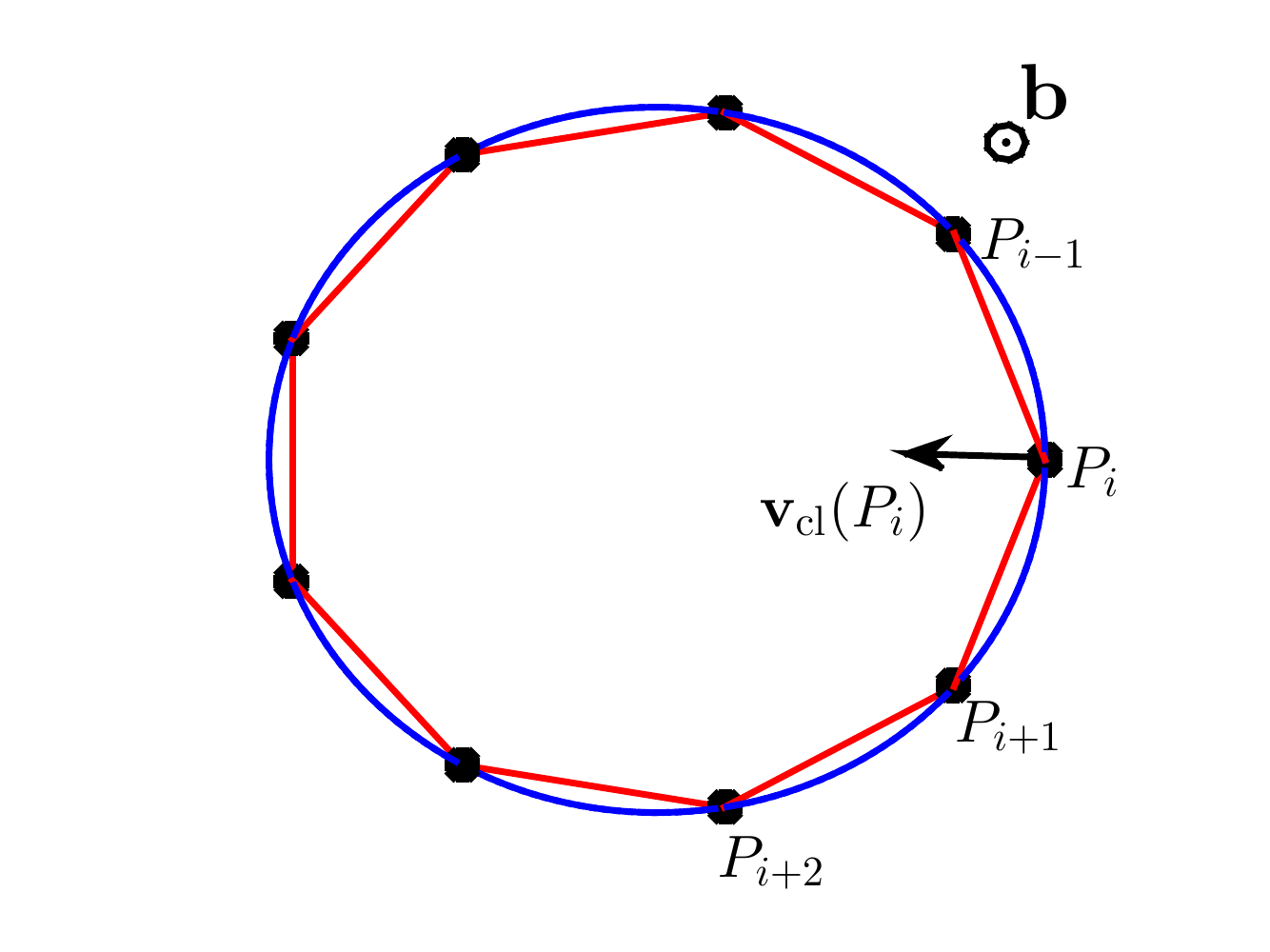}
\caption{Schematic illustration of the numerical discretization of a dislocation loop in DDD frameworks for a prismatic loop. The loop (in blue) is discretized into straight segments (in red) with $N$ representative dislocation nodes $P_i$ (in black), $i=1,2,\cdots,N$.  }
\label{fig:ddd-scha}
\end{figure}

 In this section, we present a numerical implementation method of the formulation of the self-climb velocity given in Eq.~\eqref{eq:climb-v} (or \eqref{eq:climb-v0}) in DDD simulations. Recall that in  DDD simulation methods, e.g. \cite{Kubin,Fivel1998,Zbib,Ghoniem,Weygand2002,WangGhoniem2004,cai2006,
Arsenlis,Mordehai,Bako,Zhao2010,Gu2015},  dislocations are commonly discretized into small segments (straight or curved) by nodes on them, as illustrated in Fig.~\ref{fig:ddd-scha}.
During the simulations, the location of each node is commonly updated by the scheme
\begin{equation}
\mathbf P_i^{(n+1)}=\mathbf P_i^{(n)}+\mathbf v(\mathbf P_i^{(n)})\Delta t_n
\end{equation}
 from time $t_n$ to $t_{n+1}$, where $\mathbf P_i^{(n)}$ and $\mathbf P_i^{(n+1)}$ are the locations of the point $\mathbf P_i$ at time $t_n$ and $t_{n+1}$, respectively, and $\Delta t_n=t_{n+1}-t_n$ is the time step.  The velocity $\mathbf v$ may be the glide velocity $\mathbf{v}_g$ or the climb velocity $\mathbf v_{\rm cl}^{\rm (b)}$ or
$\mathbf v_{\rm cl}^{\rm (p)}$, or a combination of them.
The glide velocity is commonly calculated by  a  mobility law
$\mathbf{v}_g=\mathbf{M}_g\cdot\mathbf{f}_{g}$,
where $\mathbf{M}_g$ is the glide mobility tensor and $\mathbf{f}_g$ is the glide Peach-Koehler force. The climb velocity due to vacancy bulk diffusion $\mathbf v_{\rm cl}^{\rm (b)}$ at all the nodal points can be obtained accurately by solving a linear system associated with the climb force $f_{\rm cl}$ as described in Ref.~\cite{Gu2015}, or other approximation methods in the literature. The long-range Peach-Koehler force, including both the glide component $f_{\rm g}$ and the climb component $f_{\rm cl}$, can be calculated by contributions from all dislocation segments by either direct summation or more efficient methods, e.g. the fast multipole method, as described in the above references.
In this paper, the Peach Koehler forces including the climb force $f_{\rm cl}$ and the glide force $f_{\rm g}$ are calculated using the method and code in Ref.~\cite{cai2006}.

To numerically implement the dislocation self-climb
 formulated in Eq.~\eqref{eq:climb-v}, we use the following finite difference scheme along the dislocations:
 \begin{equation}\label{eqn:scheme}
v_{\rm cl}^{\rm (p)}(\mathbf P_n)=c_0D_cb \frac{(q(\mathbf P_{n+1})-q(\mathbf P_n))/\Delta s_n
 -(q(\mathbf P_{n})-g(\mathbf P_{n-1}))/\Delta s_{n-1}}{(\Delta s_{n}+\Delta s_{n-1})/2},
 \end{equation}
 where $q=e^{-\frac{f_{\rm cl}(\mathbf{P})\Omega}{bk_BT}}$ (or $q=-\frac{f_{\rm cl}(\mathbf{P})\Omega}{bk_BT}$ for Eq.~\eqref{eq:climb-v0}), and $\Delta s_i=|\mathbf P_{i+1}-\mathbf P_{i}|$ is the length of the segment between the nodes $\mathbf P_{i}$ and $\mathbf P_{i+1}$.
 This formulation can be easily incorporated in the available DDD simulation methods and codes for the motions of glide and climb assisted by vacancy bulk diffusion (e.g. those reviewed in the introduction), as an additional contribution to the dislocation velocity.

\section{Properties of self-climb of prismatic loops}\label{sec:properties}

\begin{figure}[htbp]
 \centering
 \subfigure[]
 {\includegraphics[width=7cm]{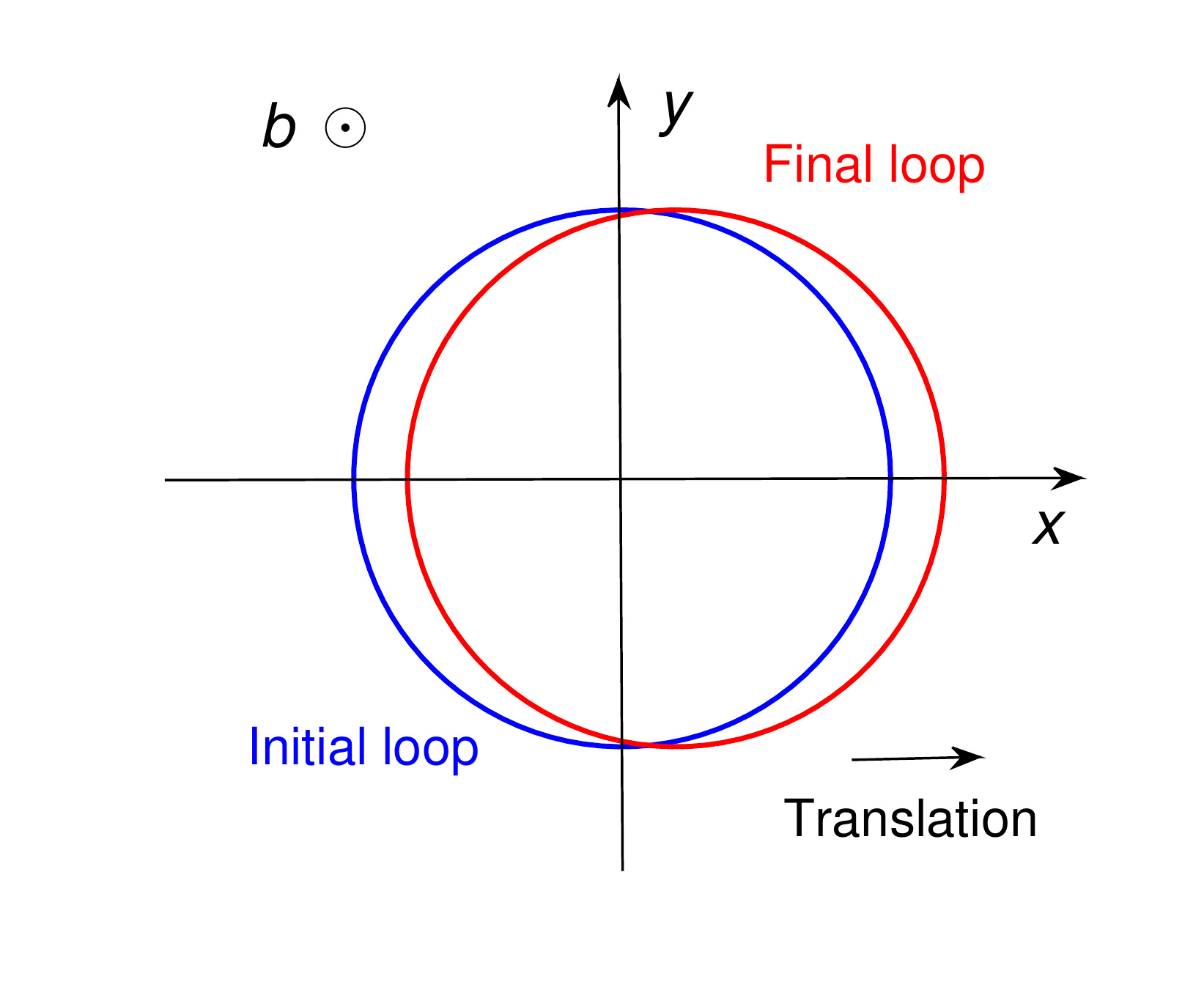}}
 \subfigure[]
 {\includegraphics[width=7cm]{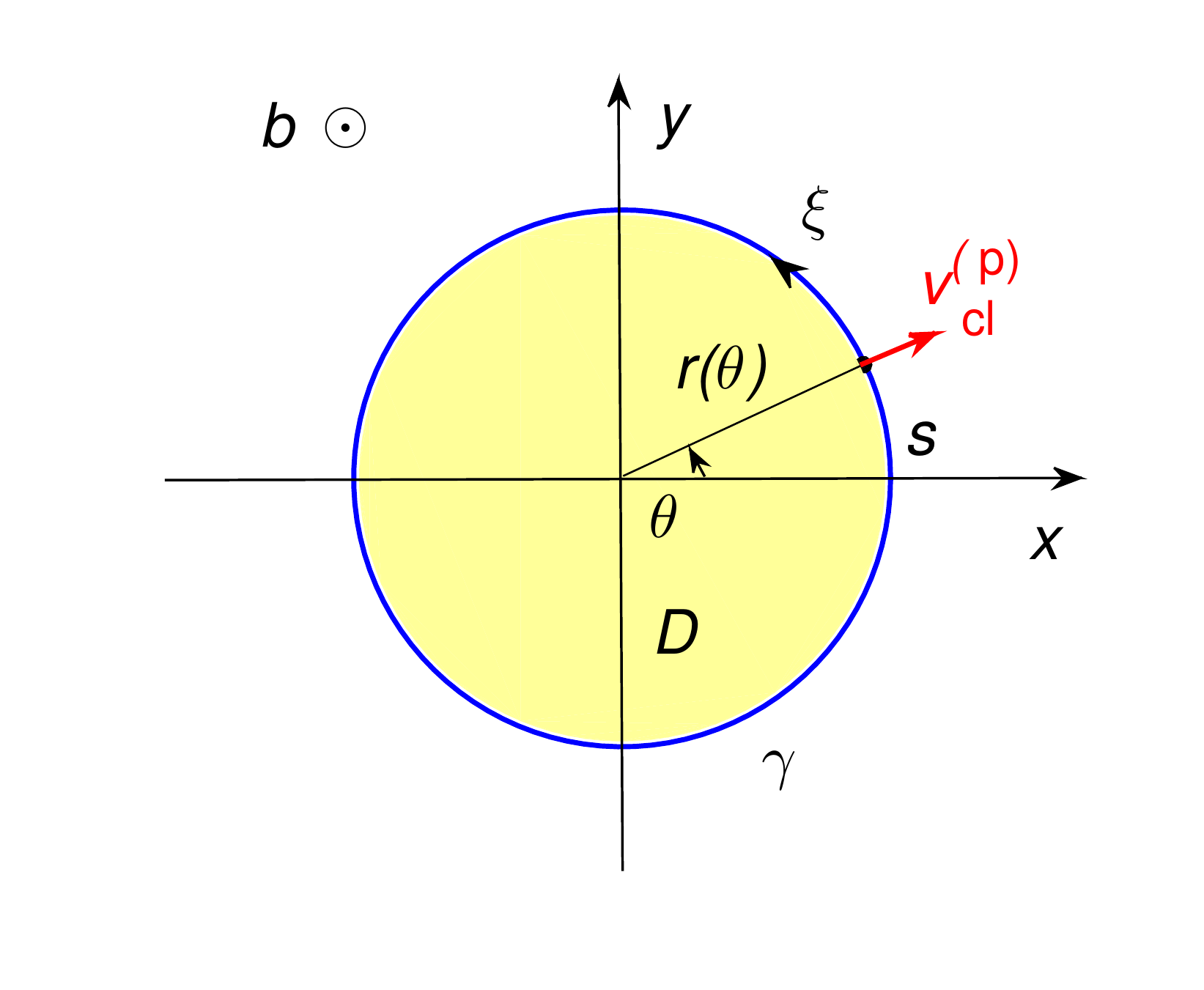}}
  \caption{(a) Translation of a prismatic loop by vacancy pipe diffusion. (b) Parametrization of a prismatic loop by angle $\theta$, $0\leq \theta\leq 2\pi$. }
  \label{fig:translation0}
\end{figure}

At low temperatures, it has been observed in experiments \cite{Kroupa1961,Hirth-Lothe} that prismatic dislocation loops can translate under a stress gradient, see an illustration in Fig.~\ref{fig:translation0}(a).
 In this case, vacancy bulk diffusion is negligible, and the dislocations climb is entirely due to
the self-climb via vacancy pipe diffusion. The area enclosed by a prismatic loop is observed to be conserved during the translation process, thus this translation by self-climb is also referred to as {\it conservative climb} of prismatic loops.
It has been shown in Ref.~\cite{Niu} that the formulation in Eq.~\eqref{eq:climb-v} is able to successfully predict the conservation of the enclosed area of a prismatic loop during the self-climb process.
In this section, we systematically study the properties of self-climb (conservative climb) of prismatic loops based on this formulation.

\subsection{Conservation of the enclosed area}\label{sec:area}

We first review the property shown in Ref.~\cite{Niu} that following the formulation of dislocation self-climb given by Eq.~\eqref{eq:climb-v},
the area enclosed by a prismatic loop is conserved.

 Consider a prismatic dislocation loop $\gamma$ moving with the  self-climb velocity in Eq.~\eqref{eq:climb-v}, the rate of change of the area $S$ enclosed by the loop is
\begin{eqnarray}\label{eqn:climb-dv1}
\frac{\D S}{\D t}=\int_\gamma v_{\rm cl}^{(\rm p)}\D s
=c_0D_cb\int_\gamma  \frac{\D^2 }{\D s^2}\left(e^{-\frac{f_{\rm cl}\Omega}{bk_BT}}\right)\D s=0.
\end{eqnarray}
This shows that using our formulation, the area $S$ that is enclosed by the loop is unchanged during the self-climb motion, which agrees with the experimental observations~\cite{Kroupa1961,Hirth-Lothe} for the motion of small prismatic loops by self-climb. In fact, by our self-climb formulation, the area enclosed by a prismatic loop is always conserved by the self-climb motion, for any size and shape of the loop, although the shape of the loop may change.

Note that the self-climb velocity $v_{\rm cl}^{(\rm p)}$ is in the normal direction of the prismatic loop. For a vacancy loop, $v_{\rm cl}^{(\rm p)}$ is in the outer normal direction, and an interstitial loop, inner normal direction. For an interstitial loop, there will be a negative sign before  $v_{\rm cl}^{(\rm p)}$ in the above proof, and the result remains $0$.

\subsection{Equilibrium shape}\label{sec:shape}

Following  Eq.~\eqref{eq:climb-v}, the equilibrium shape of a prismatic loop under self-climb satisfies $v_{\rm cl}^{(\rm p)}=0$, which leads to constant $e^{-\frac{f_{\rm cl}(\mathbf{x})\Omega}{bk_BT}}$ along the loop, or
\begin{equation}
f_{\rm cl}={\rm constant}
\end{equation}
along the loop. This shows that the equilibrium shape of a prismatic loop under self-climb is a circle when only the self-stress of the loop is considered. The equilibrium shape of a prismatic loop under self-climb is still a circle with constant stress field in addition to its self-stress.

We will further discuss properties of circular prismatic loops by self-climb motion in the following subsections. Note that the actual shape of a dislocation loop during the self-climb motion depends on the nature of the total stress field.

\subsection{Circular loop preserving}\label{circular}

In this subsection, we show that a circular prismatic loop will remain circular by self-climb under constant stress gradient, using the velocity formulation in Eq.~\eqref{eq:climb-v}.

 Consider a prismatic loop $\gamma$ in the $xy$ plane. Without loss of generality, we assume that the Burgers vector of the loop in the $+z$ direction and the loop is in the counterclockwise direction, i.e., the loop is a vacancy loop. The calculations are the same for a interstitial loop except that the climb direction changes to its opposite. The loop $\gamma$ can be parametrized by
\begin{eqnarray}\label{eq-circle-1}
\left\{
\begin{array}{l}
x(\theta,t)=r(\theta,t)\cos\theta\vspace{1ex}\\
y(\theta,t)=r(\theta,t)\sin\theta,
\end{array}
\right.
\end{eqnarray}
where $(x,y)$ is a point on the loop and $(r,\theta)$ is its polar coordinate, and $0\leq\theta\leq 2\pi$ serves as a parameter of the loop, see Fig.~\ref{fig:translation0}(b).
 By Eq.~\eqref{climbdirections}, the self-climb velocity $v_{\rm cl}^{\rm (p)}(\theta,t)$ is along normal direction of the loop and is positive in the outer normal direction.
Assume at the initial time $t=t_0$, the loop is circular:
\begin{equation}\label{eq-circle-0}
r(\theta,t_0)=R.
\end{equation}
The curvature at any point on this circular loop  is $\kappa(\theta,t_0)=1/R$.

After the loop evolves under the self-climb velocity $v_{\rm cl}^{\rm (p)}$ to the time $t_0+\Delta t$, where $\Delta t$ is small, it becomes
\begin{eqnarray}\label{eq-circle-2}
\left\{
\begin{array}{l}
x(\theta,t_0+\Delta t)\approx(R+v_{\rm cl}^{\rm (p)}(\theta,t_0)\Delta t)\cos\theta\vspace{1ex}\\
y(\theta,t_0+\Delta t)\approx(R+v_{\rm cl}^{\rm (p)} (\theta,t_0)\Delta t)\sin\theta,
\end{array}
\right.
\end{eqnarray}
and the curvature at any point on the loop can be calculated as
\begin{eqnarray}\label{eq-circle-3}
\kappa(\theta,t_0+\Delta t)
=\frac{\frac{\D x}{\D \theta}\frac{\D^2 y}{\D \theta^2}
-\frac{\D^2 x}{\D \theta^2}\frac{\D y}{\D \theta}}
{\left[\left(\frac{\D x}{\D \theta}\right)^2+\left(\frac{\D y}{\D \theta}\right)^2\right]^{\frac{3}{2}}}
 \approx \frac{R^2+(-R\frac{\D^2 v_{\rm cl}^{\rm (p)}}{\D \theta^2}+2Rv_{\rm cl}^{\rm (p)})\Delta t}{(R^2+2Rv_{\rm cl}^{\rm (p)} \Delta t)^{\frac{3}{2}}}.
\end{eqnarray}
The rate of change of the curvature is
\begin{eqnarray}\label{eq-circle-4}
\frac{\partial \kappa(\theta,t_0)}{\partial t}&=&\lim_{\Delta t\rightarrow0}\frac{\kappa(\theta,t_0+\Delta t)-\kappa(\theta,t_0)}{\Delta t} = -\frac{\frac{\D^2 v_{\rm cl}^{\rm (p)}}{\D \theta^2}+v_{\rm cl}^{\rm (p)}}{R^2}.
\end{eqnarray}
The curvature is preserved when $\frac{\partial \kappa(\theta,t_0)}{\partial t}=0$, or
 \begin{eqnarray}\label{eq-circle-conservative-condition}
\frac{\D^2 v_{\rm cl}^{\rm (p)}}{\D \theta^2}+v_{\rm cl}^{\rm (p)}=0.
\end{eqnarray}
This is equivalent to
\begin{equation}
v_{\rm cl}^{\rm (p)}=A_v\cos\theta+B_v\sin\theta,
\end{equation}
for constants $A_v$ and $B_v$ independent of $\theta$.

Using the self-climb velocity in Eq.~\eqref{eq:climb-v}, after integration twice, the above circular loop preserving conditions becomes
$e^{-\frac{f_{\rm cl}\Omega}{bk_BT}}=A_f\cos\theta+B_f\sin\theta+C_f$, where $A_f$, $B_f$ and $C_f$ are some constants.
When $f_{\rm cl}\Omega<<bk_BT$, we have $e^{-\frac{f_{\rm cl}\Omega}{bk_BT}}\approx 1-\frac{f_{\rm cl}\Omega}{bk_BT}$, the circular loop preserving condition becomes
\begin{eqnarray}\label{eq-circle-conservative-condition-f1}
f_{\rm cl}=A\cos\theta+B\sin\theta+C,
\end{eqnarray}
for some constant coefficients $A$, $B$ and $C$.
When this prismatic loop is place in a stress field with constant normal stress gradient, i.e.,
\begin{eqnarray}\label{constant-gradient}
\sigma_{33}=\frac{\partial \sigma_{33}}{\partial x}x+\frac{\partial \sigma_{33}}{\partial y}y+\sigma_{33}^0,
 \end{eqnarray}
 where $\frac{\partial \sigma_{33}}{\partial x}$, $\frac{\partial \sigma_{33}}{\partial y}$ and $\sigma_{33}^0$ are constants,
 on the circular loop described by Eqs.~\eqref{eq-circle-1} and \eqref{eq-circle-0}, using Eq.~\eqref{eqn:fcl}, we have
\begin{eqnarray}\label{eq-circle-conservative-condition-f2}
f_{\rm cl}=-\sigma_{33}^{\rm total}b=-\frac{\partial \sigma_{33}}{\partial x}bR\cos\theta
-\frac{\partial \sigma_{33}}{\partial y}bR\sin\theta-\sigma_{33}^0b-\sigma_{33}^{\rm self}b.
\end{eqnarray}
Here $\sigma_{33}^{\rm total}=\sigma_{33}+\sigma_{33}^{\rm self}$, with the self stress $\sigma_{33}^{\rm self}$ of this circular prismatic loop being constant along the loop. Thus the circular loop preserving condition in Eq.~\eqref{eq-circle-conservative-condition-f1} is satisfied. Note that the self-stress generated by the circular loop itself is constant along the loop.
 Recall that as shown in Sec.~\ref{sec:area}, the area enclosed by the loop is preserved. Therefore, we can conclude that a circular prismatic loop will translate rigidly with circular shape preserved under constant stress gradient by self-climb.

 Note that when the stress gradient varies slowly, it can be locally considered as constant, thus a small circular prismatic loop will always translate rigidly with circular shape preserved. This explains the classical experimental observations \cite{Kroupa1961,Hirth-Lothe}.

 \subsection{Translation velocity}\label{avelocity}

In this subsection, we calculate the average translation velocity of a prismatic loop  by self-climb under stress gradient.

Consider a prismatic loop $\gamma$, and the region enclosed by this loop is denoted by $D$, see Fig.~\ref{fig:translation0}(b) for an example. As in the previous subsection, without loss of generality, we assume that the Burgers vector of the loop is in the $+z$ direction and the loop is in the counterclockwise direction.
The average velocity of $\gamma$ can be calculated by the velocity of the mass center of the region $D$ enclosed by it, which is
\begin{eqnarray}
\bar{\mathbf{v}}=\frac{\D}{\D t}\left(\frac{\iint_D(x, y)\D S }{\iint_D \D S}\right),
\end{eqnarray}
where $\D S$ is the area element.

As already shown in Sec.~\ref{sec:area},
 the area of the region enclosed by $\gamma$, or $\iint_D \D S$, remains the same during the translation of $\gamma$, further using the transport theorem \cite{Fung}, we have
\begin{eqnarray}\label{centervelocity}
\bar{\mathbf{v}}=\frac{1}{\iint_D \D S}\left(\frac{\D}{\D t}\iint_D x \D S,  \frac{\D}{\D t}\iint_D y \D S\right)=\frac{1}{\iint_D \D S}\left(\int_\gamma xv_{\rm cl}^{\rm (p)} \D s,  \int_\gamma y v_{\rm cl}^{\rm (p)}\D s\right),
\end{eqnarray}
where the local self-climb velocity $v_{\rm cl}^{\rm (p)}$ is given  in Eq.~\eqref{eq:climb-v}. (Recall that $v_{\rm cl}^{\rm (p)}$ is in the outer normal direction of this vacancy loop.)
This average velocity formula applies to a prismatic loop with any shape and in any stress field.

This translation velocity has an analytical formula when the prismatic loop $\gamma$ is circular and the stress gradient is constant. Recall that a circular prismatic loop
will remain circular during the self-climb motion as shown in Sec.~\ref{circular}. In this case, the loop $\gamma$ is parameterized as $x=R\cos \theta+x_0$, $y= R\sin \theta+y_0$, $0\leq\theta\leq 2\pi$, and $\D s=R\D \theta$ in Eq.~\eqref{centervelocity};  the climb force $f_{\rm cl}$ takes the form in Eq.~\eqref{eq-circle-conservative-condition-f2}.
Substituting the self-climb velocity formula in Eq.~\eqref{eq:climb-v} into Eq.~\eqref{centervelocity},
and performing integration by parts twice, we have
\begin{eqnarray}\label{eq-average-velocity} \nonumber
\bar{\mathbf{v}}&=&
-\frac{c_0D_cb}{\pi R^2}\left(\int_0^{2\pi}  e^{-\frac{f_{\rm cl}\Omega}{bk_BT}} \cos \theta \D\theta, \int_0^{2\pi}  e^{-\frac{f_{\rm cl}\Omega}{bk_BT}}  \sin \theta \D\theta\right)\\
&=&-\frac{c_0D_cb}{\pi R^2}e^{\frac{\left(\sigma_{33}^0+\sigma_{33}^{\rm self}\right)\Omega}{k_BT}} \left(\int_0^{2\pi}  e^{\frac{\Omega R}{k_BT}\frac{\partial \sigma_{33}}{\partial x} \cos\theta} \cos \theta \D \theta, \int_0^{2\pi}  e^{\frac{\Omega R}{k_BT}\frac{\partial \sigma_{33}}{\partial y}\sin\theta}      \sin \theta \D\theta\right)\nonumber \\
&=&-\frac{2c_0D_ce^{\frac{\left(\sigma_{33}^0+\sigma_{33}^{\rm self}\right)\Omega}{k_BT}}b}{R^2}
\left(I_1\left({\textstyle \frac{\Omega R}{k_BT}\frac{\partial \sigma_{33}}{\partial x}}\right), I_1\left({\textstyle \frac{\Omega R}{k_BT}\frac{\partial \sigma_{33}}{\partial y}}\right)\right),
\end{eqnarray}
where $I_1(\cdot)$ is a modified Bessel function of the first kind. Recall that $\sigma_{33}^0$ is the constant  stress (excluding the self-stress) and $\sigma_{33}^{\rm self}$ is the constant self-stress on the circular loop.

When the stress gradient is small or the radius of the loop is small, i.e. $R|\nabla \sigma_{33}|<<k_BT/\Omega$, using the property of $I_1(z)\approx z/2$ for small $z$, the translation velocity of the circular prismatic loop can be simplified as
\begin{eqnarray}\label{eq-average-velocity0}
\bar{\mathbf{v}}=-\frac{c_0 D_c e^{\frac{\left(\sigma_{33}^0+\sigma_{33}^{\rm self}\right)\Omega}{k_BT}}\Omega b}{k_BTR}
\left(\frac{\partial \sigma_{33}}{\partial x}, \frac{\partial \sigma_{33}}{\partial y}\right)
=-\frac{c_0 D_ce^{\frac{\left(\sigma_{33}^0+\sigma_{33}^{\rm self}\right)\Omega}{k_BT}}\Omega b}{k_BTR}\nabla\sigma_{33}.
\end{eqnarray}
 This means that when the external stress field is a linear function of the space coordinates, the translation velocity of a circular prismatic loop by self-climb is linearly proportional to the constant stress gradient vector and is in the same direction with it. Eq.~\eqref{eq-average-velocity0} also applies to small circular prismatic loops with a general external stress field, because in this case, the stress gradient is approximately constant near the small prismatic loop.

For an interstitial loop, the translation velocity formulas in Eqs.~\eqref{eq-average-velocity} and \eqref{eq-average-velocity0} will change their sign, and become
$\bar{\mathbf{v}}=\frac{2c_0D_ce^{\frac{(\sigma_{33}^0+\sigma_{33}^{\rm self})\Omega}{k_BT}}b}{R^2}
\left(I_1\left(\frac{\Omega R}{k_BT}\frac{\partial \sigma_{33}}{\partial x}\right), I_1\left(\frac{\Omega R}{k_BT}\frac{\partial \sigma_{33}}{\partial y}\right)\right)$
and
$\bar{\mathbf{v}}=\frac{c_0 D_ce^{\frac{\left(\sigma_{33}^0+\sigma_{33}^{\rm self}\right)\Omega}{k_BT}}\Omega b}{k_BTR} \nabla\sigma_{33}$, respectively.

 Note that this translation velocity formula of a circular prismatic loop in  Eq.~\eqref{eq-average-velocity} (or its simplified form in  Eq.~\eqref{eq-average-velocity0}) is derived from the dislocation dynamics formulation for self-climb in Eq.~\eqref{eq:climb-v}, which specifies the self-climb velocity of each point on the dislocation loop. The available theories in the literature for the self-climb of prismatic loops \cite{Johnson1960,Turnbull1970,Narayan1972,Swinburne2016,Okita2016} are all based on the assumption of rigid translation of a circular prismatic loop, and their translation velocities are expressed in terms of the driving force and mobility of the entire loop.  For example, the translation velocity of a circular prismatic loop with radius $R$ is $\mathbf{v}=-\frac{2\nu_a e^{-\frac{E_{\rm sc}}{k_BT}}a^5}{\pi k_BTR^3}\nabla E$ in Ref.~\cite{Swinburne2016}, where $a$ is the lattice constant, $\nu_a$ is an attempt frequency, $E_{\rm sc}$ is a characteristic activation energy for the self-climb, and $E$ is the interaction energy between the prismatic loop and the stress field (excluding the self-stress of the loop). Now we include a comparison of our formulation with those available theories~\cite{hkust}.
 In fact, the interaction energy $E$
  can be calculated as $E=\iint_D b\sigma_{33}\D S$ \cite{Hirth-Lothe}. Using the parameterization of the circular loop given above, we can calculate that $\nabla E=\pi R^2 b\nabla\sigma_{33}$ when $\nabla\sigma_{33}$ is a constant vector. The translation velocity in Eq.~\eqref{eq-average-velocity0} can then be written as
  \begin{equation}
\bar{\mathbf{v}}=-\frac{c_0D_ce^{\frac{\left(\sigma_{33}^0+\sigma_{33}^{\rm self}\right)\Omega}{k_BT}}\Omega}{\pi k_BTR^3}\nabla E,
\end{equation}
 which is in excellent agreement with the available theories.

The prefactor $c_0e^{\frac{\left(\sigma_{33}^0+\sigma_{33}^{\rm self}\right)\Omega}{k_BT}}D_c$ in our formulation is based on the vacancy pipe diffusion mechanism of the self-climb. Recall that $c_0$ is the equilibrium vacancy concentration on the dislocation loop and $D_c$ is the vacancy pipe diffusion constant. The product $c_0e^{\frac{\left(\sigma_{33}^0+\sigma_{33}^{\rm self}\right)\Omega}{k_BT}}$ is the stress-dependent vacancy concentration on the dislocation loop.
It has been shown by experiments and atomistic simulations that $D_c$ is much larger than $D_v$ (vacancy diffusion constant in the bulk)~\cite{Hirth-Lothe,Balluffi1971,Miller1981,Hoagland1998,Fang2000,Mishin2009}.
On the other hand, the prefactor $2\nu_a e^{-\frac{E_{\rm sc}}{k_BT}}a^5$ in the available theories in the literature is based on the attempt frequency and energy barrier for rigid translation of the prismatic loop.  In Ref.~\cite{Swinburne2016}, it was shown by kinetic Monte Carlo simulations that the  characteristic activation energy barrier for self-climb $E_{\rm sc}$ is much lower than the effective activation energy of vacancy bulk diffusion in bcc metals. Both types of models describe fast vacancy pipe diffusion than vacancy bulk diffusion in the self-climb process.


 Our discussion above also shows that the linear relationship between the translation velocity and the energy gradient $\nabla E$ widely adopted in the literature only holds when the stress gradient is small or the radius of the loop is small, otherwise, the more accurate translation velocity formula in  Eq.~\eqref{eq-average-velocity} should be used. When the prismatic loop is not circular, its pointwise self-climb velocity can be calculated by Eq.~\eqref{eq:climb-v0} or \eqref{eq:climb-v}.


\section{DDD Simulations}\label{sec:simulations}

In this section, we perform  DDD simulations of the self-climb of dislocations  and compare the results with experimental observations and atomistic simulations. The numerical scheme is given in Eq.~\eqref{eqn:scheme} which is a discretization of the formulation in Eq.~\eqref{eq:climb-v}. (The simulation results will be similar by using the simplified formulation in Eq.~\eqref{eq:climb-v0}.) Each loop is discretized into $40-60$ small segments.
 We consider the self-climb of dislocation loops in Fe with Burgers vectors $\mathbf b=a/2<111>$, which is in the $+z$ direction in the coordinate system. The lattice constant of Fe is $a=2.856\AA$. The atomic volume of Fe is $\Omega=a^3/2$, and the elastic constants are $\mu=86$GPa and $\nu=0.291$. The dislocation core parameter is $r_d=2b$. The temperature is $T=750$K. The time is rescaled in the unit of $b^2/(c_0D_c)$.

\subsection{Conservation of enclosed area}

We have shown in Ref.~\cite{Niu} and reviewed in Sec.~\ref{sec:area} that following the formulation of dislocation self-climb given by Eq.~\eqref{eq:climb-v} or \eqref{eq:climb-v0}, the area enclosed by a prismatic loop is conserved. Here we present a numerical examination.

\begin{figure}[!htbp]
\centering
\includegraphics[width=12cm]{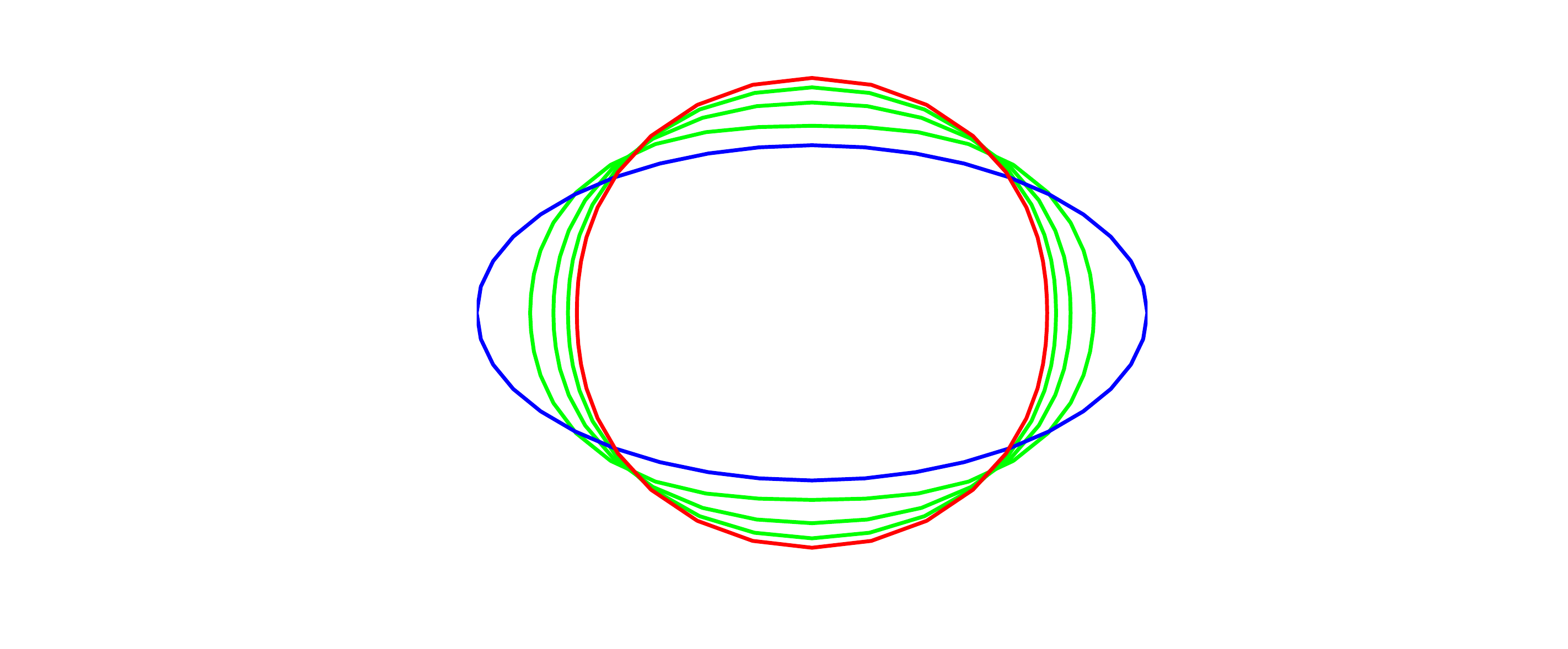}
\caption{Numerical simulation of self-climb of an elliptic prismatic loop (in blue), which converges to a circular loop (in red). Some snapshots of the loop in equal time intervals during the evolution are also shown (in green). }
\label{fig:ddd-schb}
\end{figure}

We start from a prismatic loop in the $xy$ plane with ellipse shape, whose two axes are $l_1=80b$ and $l_2=40b$. The loop is in the counterclockwise direction, which means a vacancy loop. As shown in Fig.~\ref{fig:ddd-schb}, the loop evolves to a circle which is stable with further evolution. This simulation result can be understood by Eq.~\eqref{eq:climb-v} that when the prismatic loop is not circular, the self stress and accordingly the climb force $f_{\rm cl}$ is not uniform on the loop, and this drives the loop to self-climb until a constant climb force $f_{\rm cl}$ is reached, which is a circle. The radius of the stable circular loop measured from the simulation result is  $R=56.1b$. The theoretical value of radius of the circle loop under conservation of the enclosed area is  $R=\sqrt{l_1l_2}=56.6b$. Excellent agreement can be seen from the numerical simulation and theoretical prediction that the area enclosed by a prismatic loop is conserved during the self-climb motion and the loop converges to the equilibrium shape of a circle under its self-stress.  These results also explain the early experimental observations of the self-climb behaviors of prismatic loops, termed as {\it conservative climb} in some references \cite{Hirth-Lothe,Kroupa1961}.

\subsection{Translation of a circular prismatic loop under constant stress gradient}\label{translation_numerical}

In this subsection, we present a simulation of translation of a circular prismatic dislocation loop under constant stress gradient.  The initial circular prismatic loop is located in the $xy$ plane and is in the counterclockwise direction, i.e., it is a vacancy loop. The radius of the loop is $R=100b$.   The applied stress field is $\sigma_{33}^{\rm app}=-px$ with $p=10^{-5}\mu/b$, i.e., the applied stress gradient is constant and is in the $-x$ direction.

\begin{figure}[htbp]
 \centering
 \includegraphics[width=14cm]{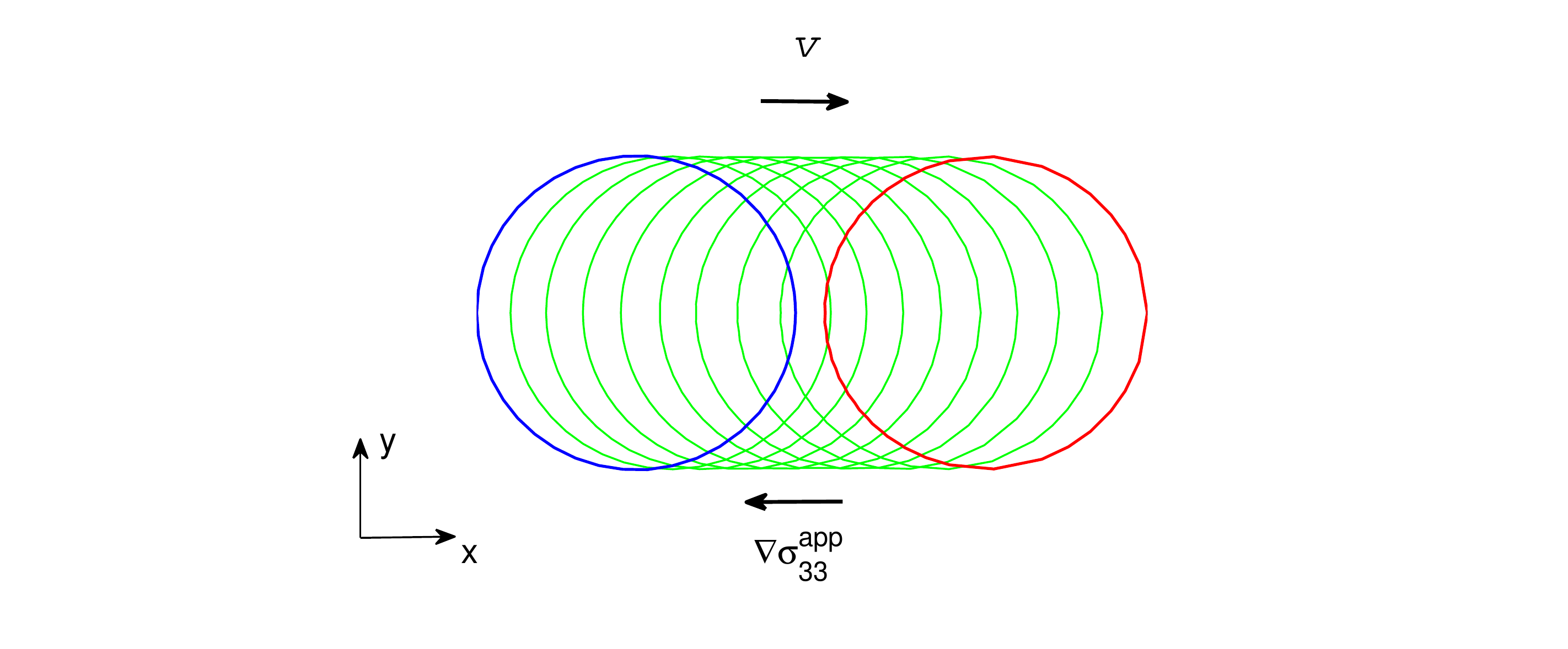}
  \caption{Simulation of translation of a circular prismatic loop under constant stress gradient.  Snapshots of the loop at $t=n\Delta t$, $n=0,1,\cdots,N$, are shown. The blue circle is  the initial loop, and the red circle is the loop at $N\Delta t$.}
  \label{fig:translationc}
\end{figure}

The simulation result is shown in Fig.~\ref{fig:translationc}.
The simulation shows that under the constant stress gradient field, the circular prismatic loop moves with a constant translation velocity while remaining  its circular shape. This behavior of translation of prismatic loops under stress gradient agrees with the experimental observations \cite{Hirth-Lothe,Kroupa1961,Turnbull1970,Narayan1972,Swinburne2016} and atomistic simulations \cite{Swinburne2016,Okita2016}. Note that this simulation is based on the pointwise self-climb velocity formulation in Eq.~\eqref{eq:climb-v}.
The translation velocity of the loop estimated from the evolution of the first $4000$ time steps is  $v=1.94\times 10^{-5}c_0D_c/b$. This value is in excellent agreement with the theoretical value  $v=1.85\times 10^{-5}c_0D_c/b$ given by the loop translation velocity formula in Eq.~\eqref{eq-average-velocity0}, where  the self stress of the loop is $\sigma_{33}^0=6.72\times 10^{-3}\mu$  from the calculations using the method in Ref.~\cite{cai2006}.

The parameters $c_0$ (reference equilibrium vacancy concentration on the dislocation loop) and $D_c$ (vacancy pipe diffusion constant) in the self-climb formulation in Eq.~\eqref{eq-average-velocity0} can be estimated as a product by comparing the atomistic simulation results or experimentally measured data, as did in the literature  \cite{Johnson1960,Turnbull1970,Narayan1972,Swinburne2016,Okita2016}  using velocity formulas based on the assumption of rigid translation of a circular prismatic loop. The connection between our formulation and those in the literature has been discussed briefly at the end of Sec.~\ref{avelocity}.

Note that in this case, the loop-size-dependent factor $e^{\frac{\sigma_{33}^0\Omega}{k_BT}}=1.92$ in the loop translation velocity formula in Eq.~\eqref{eq-average-velocity0}. Neglecting this factor will lead to an error about $92\%$ in the translation velocity obtained directly using Eq.~\eqref{eq:climb-v} by simulation.

\subsection{Coalescence of prismatic loops by self-climb}

In this subsection, we use our DDD self-climb formulation to simulate the detailed process of the coalescence of prismatic loops by self-climb that has been observed in experiments \cite{Silcox1960,Washburn1960,Johnson1960,Turnbull1970,Narayan1972,Swinburne2016}.

\begin{figure}[htbp]
\centering
\subfigure[]{\includegraphics[width=3.in]{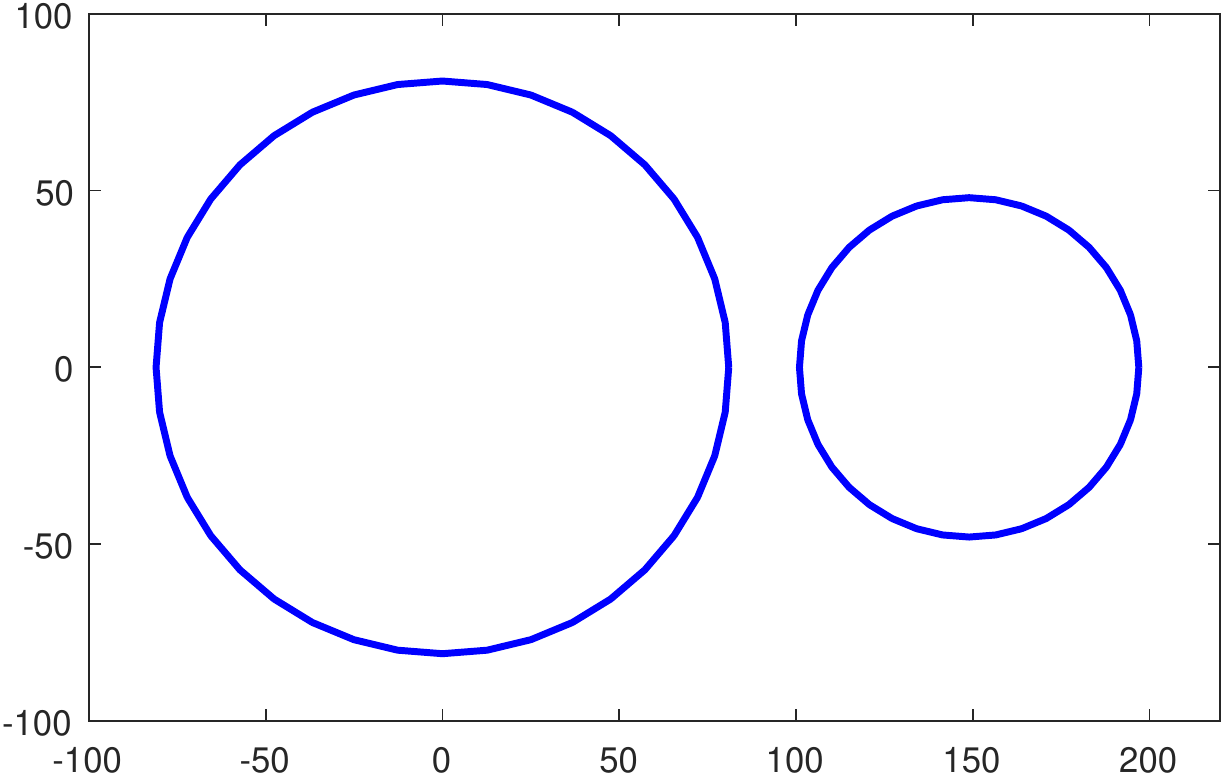}}
\subfigure[]{\includegraphics[width=3.in]{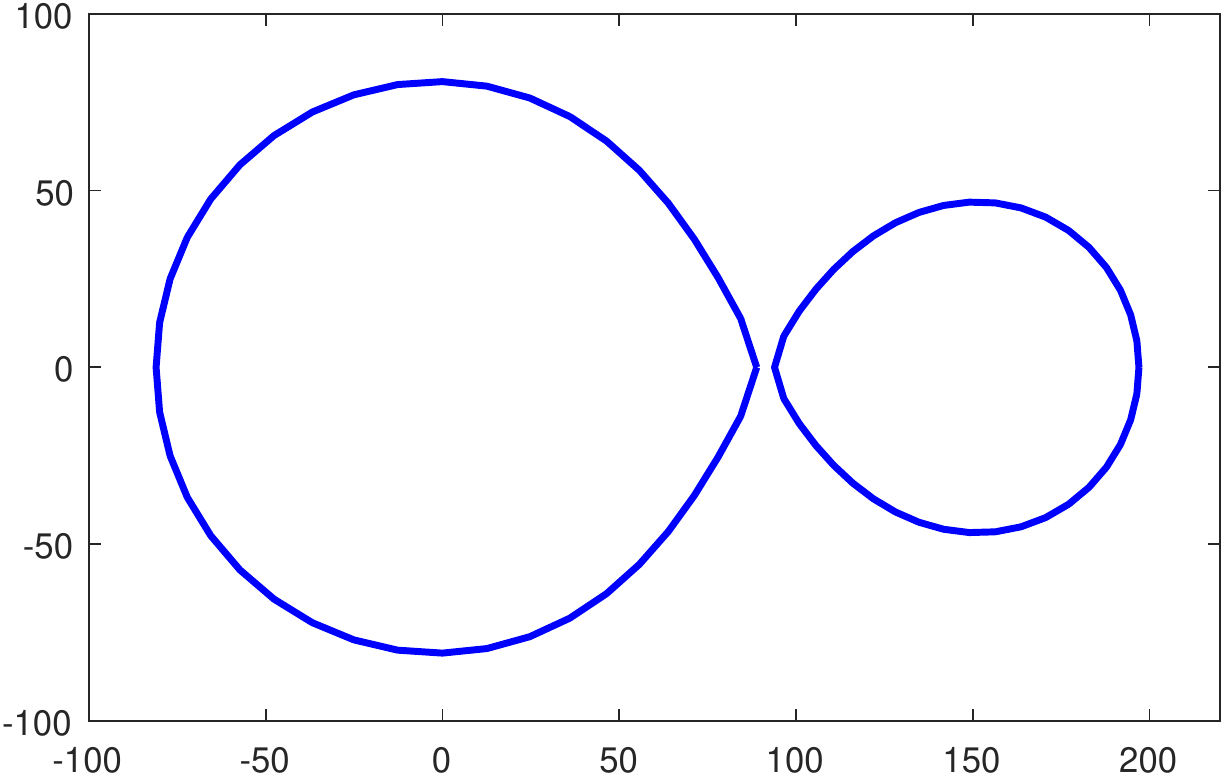}}
\subfigure[]{\includegraphics[width=3.in]{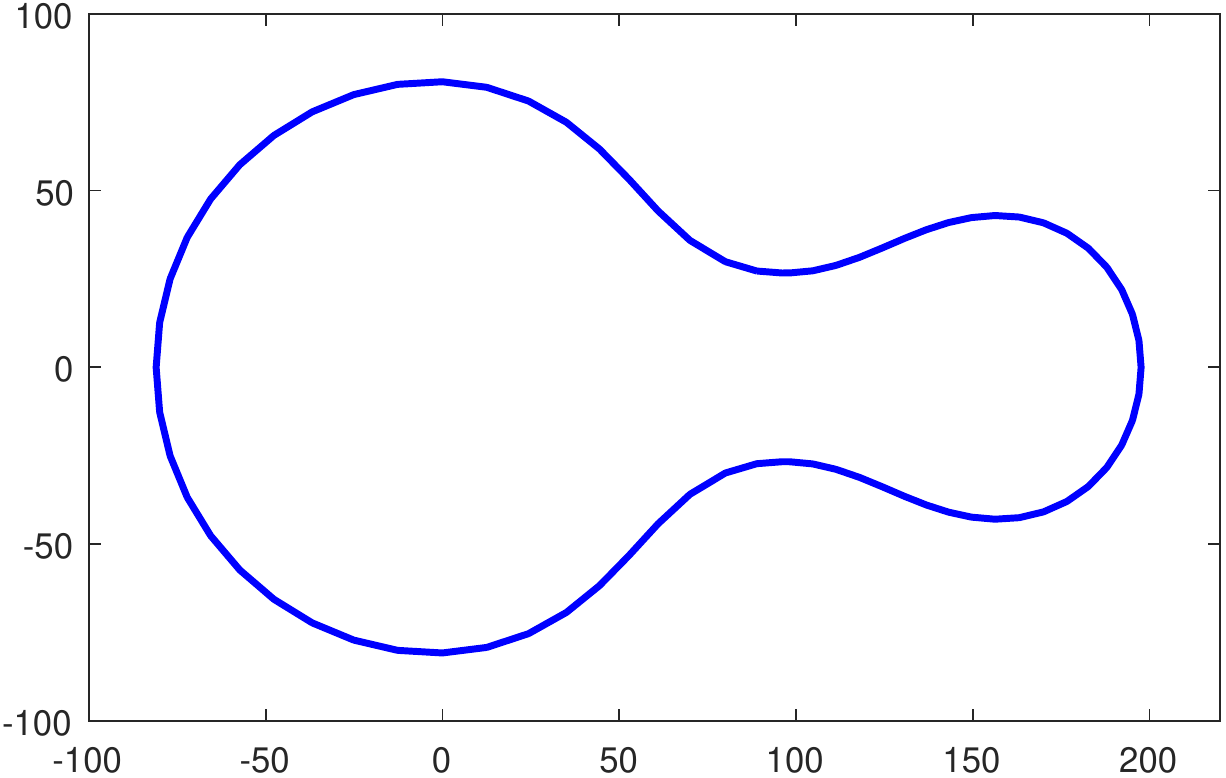}}
\subfigure[]{\includegraphics[width=3.in]{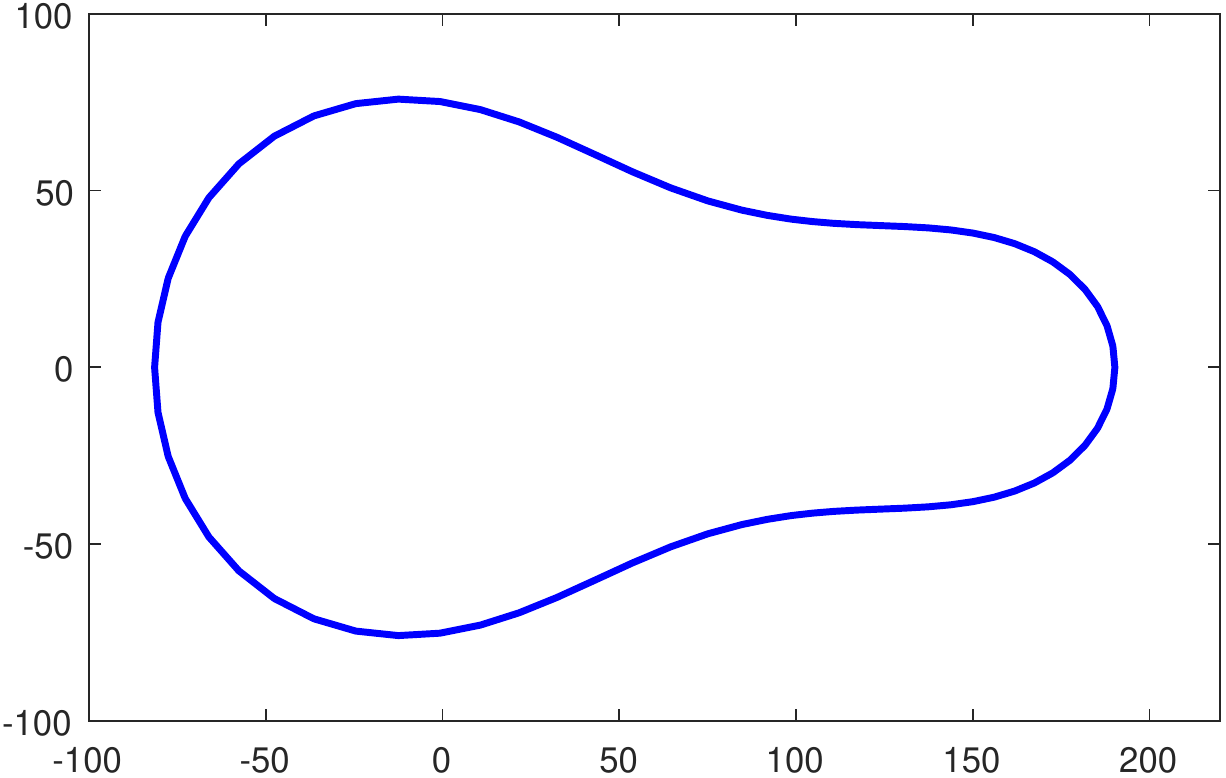}}
\subfigure[]{\includegraphics[width=3.in]{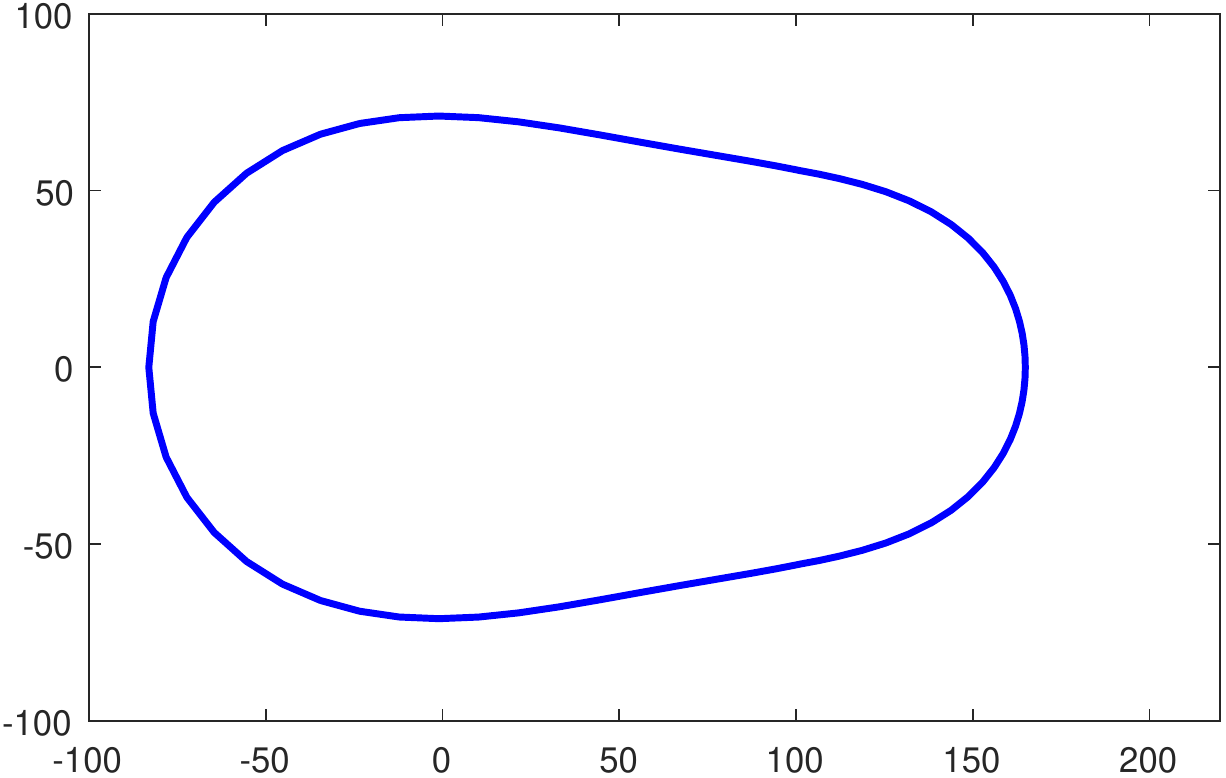}}
\subfigure[]{\includegraphics[width=3.in]{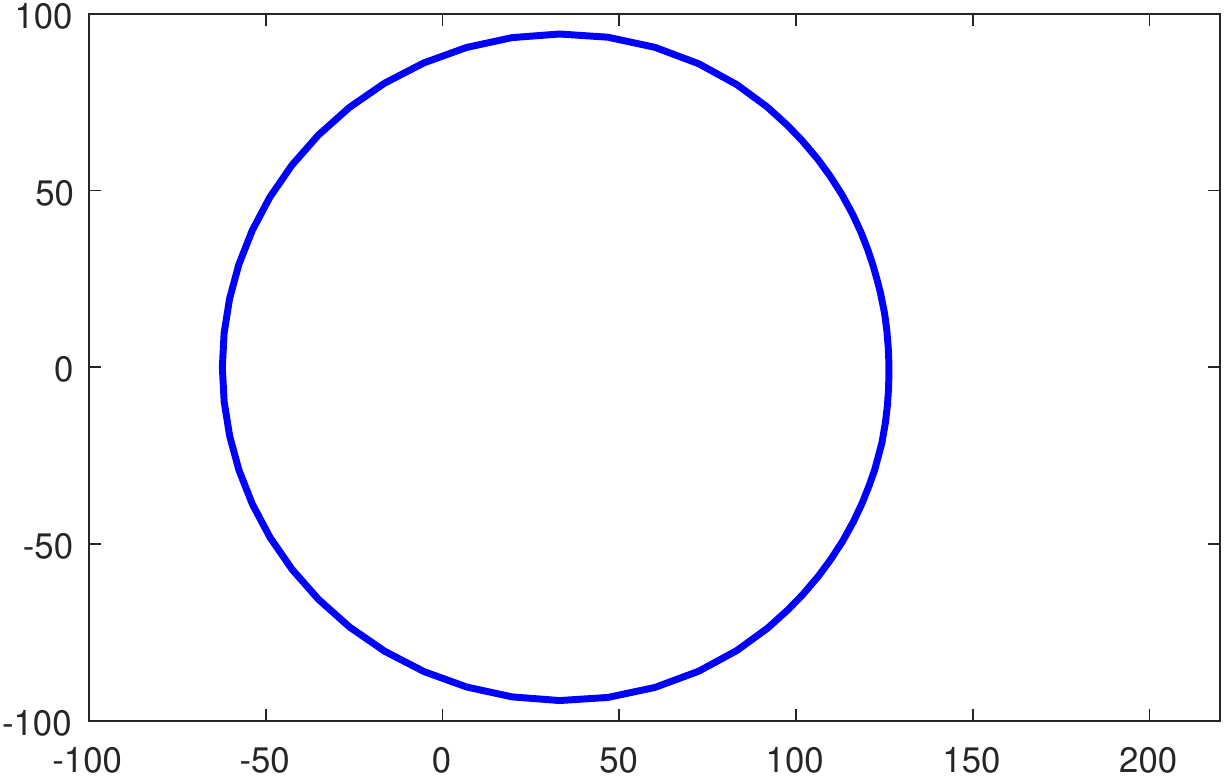}}
	\caption{Coalescence of two interstitial loops by self-climb under their elastic interaction obtained by our DDD simulation. The length unit is $b$.  These images show the simulation results at time
$t=0$s, $0.47$s, $0.49$s,  $0.70$s, $1.04$s,and  $4.10$s,
 respectively.}\label{fig.twoloopsattractive1}
\end{figure}

 We consider two circular self interstitial loops in the $xy$ plane, which are in the clockwise direction. The two loop have radii of $R_1=81b$ and $R_2=48b$, respectively, and are separated (from center to center) by $d=149b$ as shown in Fig.~\ref{fig.twoloopsattractive1}(a).  We use the same values of parameters as those for the coalescence of interstitial prismatic loops with Burgers vector $1/2[111]$ in iron at $750$K reported in Ref.~\cite{Swinburne2016} by using atomistic simulation and TEM observation. As discussed at the end of Sec.~\ref{avelocity}, we have $c_0D_c\Omega=2\nu_aa^5e^{-\frac{E_{\rm sc}}{k_BT}}$, neglecting the stress-dependence of $c_0$. Further using $\Omega=a^3/2$ in bcc Fe \cite{Swinburne2016}, we have $c_0D_c=4\nu_aa^2e^{-\frac{E_{\rm sc}}{k_BT}}$. The value of $E_{\rm sc}$ obtained in Ref.~\cite{Swinburne2016} by atomistic model is $1.34$eV and the value of $\nu_a$ used therein is $10^{13}$/s. These values of parameters are summarized in Table~\ref{table1}.

\begin{table}[htbp]
        \centering
\begin{tabular}{|c|c|c|c|c|c|c|c|c|}
  \hline
     $R_1$  &$ R_2$  &$d$    &$T$                             &$b=\sqrt{3}/2a$       &$\nu_a$              &$ E_{\rm sc}$ &$\Omega=a^3/2$   &$c_0D_c=4\nu_aa^2e^{-\frac{E_{\rm sc}}{k_BT}}$ \\
     \hline
     $20$nm   & $12$nm   &$37$nm   &$750$K         &$0.248$nm   &$10^{13}$/s    &$1.34$eV &$0.0117$nm$^3$ & $3.24\times 10^3$nm$^2$/s\\
  \hline
\end{tabular}
\caption{Values of parameters used in the simulation of coalescence of two prismatic loops in Fe with Burgers vector $1/2[111]$. These values are taken from Ref.~\cite{Swinburne2016} in their atomistic simulation and TEM observation. Values of parameters of Fe used are $\mu=86$GPa, $\nu=0.291$, and $a=0.286$nm.  }
\label{table1}
\end{table}

The detailed coalescence process obtained by our simulation is shown in Fig.~\ref{fig.twoloopsattractive1}. The two loops are attracted to each other by self-climb under the elastic interaction between them, see Fig.~\ref{fig.twoloopsattractive1}(b). The smaller loop moves faster, which is consistent with the approximate formula in Eq.~\eqref{eq-average-velocity0} and the theories in the literature \cite{Johnson1960,Turnbull1970,Narayan1972,Swinburne2016,Okita2016} for rigid translation of the loops. However, the two loops deviate from the circular shape when they are getting closed to each other. When the two loops meet, they quickly combine into a single loop, as shown in Fig.~\ref{fig.twoloopsattractive1}(c). The combined single loop eventually evolves into a stable, circular shape, see Fig.~\ref{fig.twoloopsattractive1}(d)-(e). Under conservation of the enclosed area by loops, the theoretical value of the final single circular loop has radius $R=\sqrt{R_1^2+R_2^2}=94.2b$. Numerically, the radius of the final single circular loop is measured as $R=94.3b$, which agrees perfectly with the theoretical value.

 In our DDD simulation, the time for these two circular prismatic loop to coalesce into a single circular prismatic loop is $t=2.16\times 10^5 b^2/(c_0D_c)=4.10s$. This result agrees with the coalescence times obtained in  experimental observation and atomistic simulation under the same condition reported in Ref.~\cite{Swinburne2016}, which are $30$s and $7.5$s, respectively. Note that as shown in Ref.~\cite{Swinburne2016}, using DDD simulations of climb by vacancy bulk diffusion instead of self-climb by pipe diffusion, the coalescence time is $3.3\times 10^7$s -- six orders of magnitude longer than the experimental result, indicating the dominant effect of self-climb under this condition.

\subsection{Translation of a prismatic loop by self-climb driven by a moving edge dislocation}\label{subsec:movingedge}

\begin{figure}[htbp]
\centering
\subfigure[]{\includegraphics[width=1in]{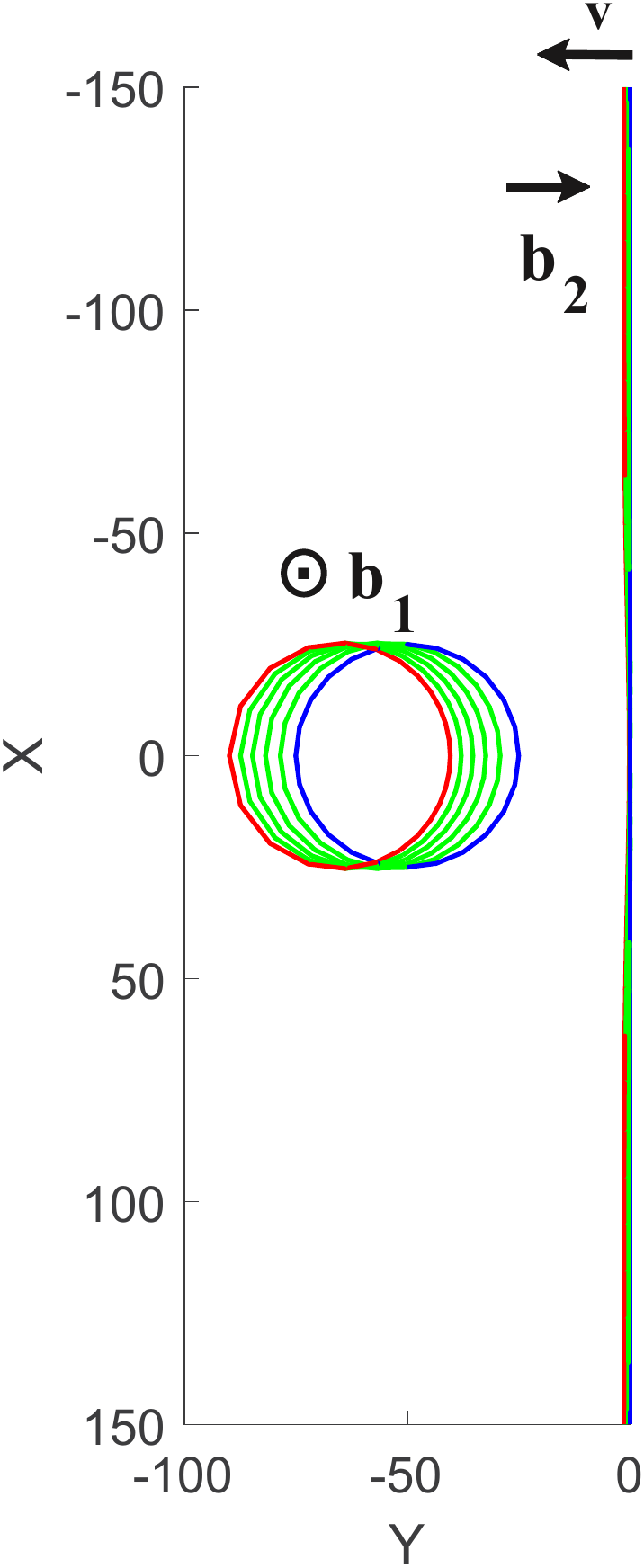}}  \hspace{1in}
\subfigure[]{\includegraphics[width=2.4in]{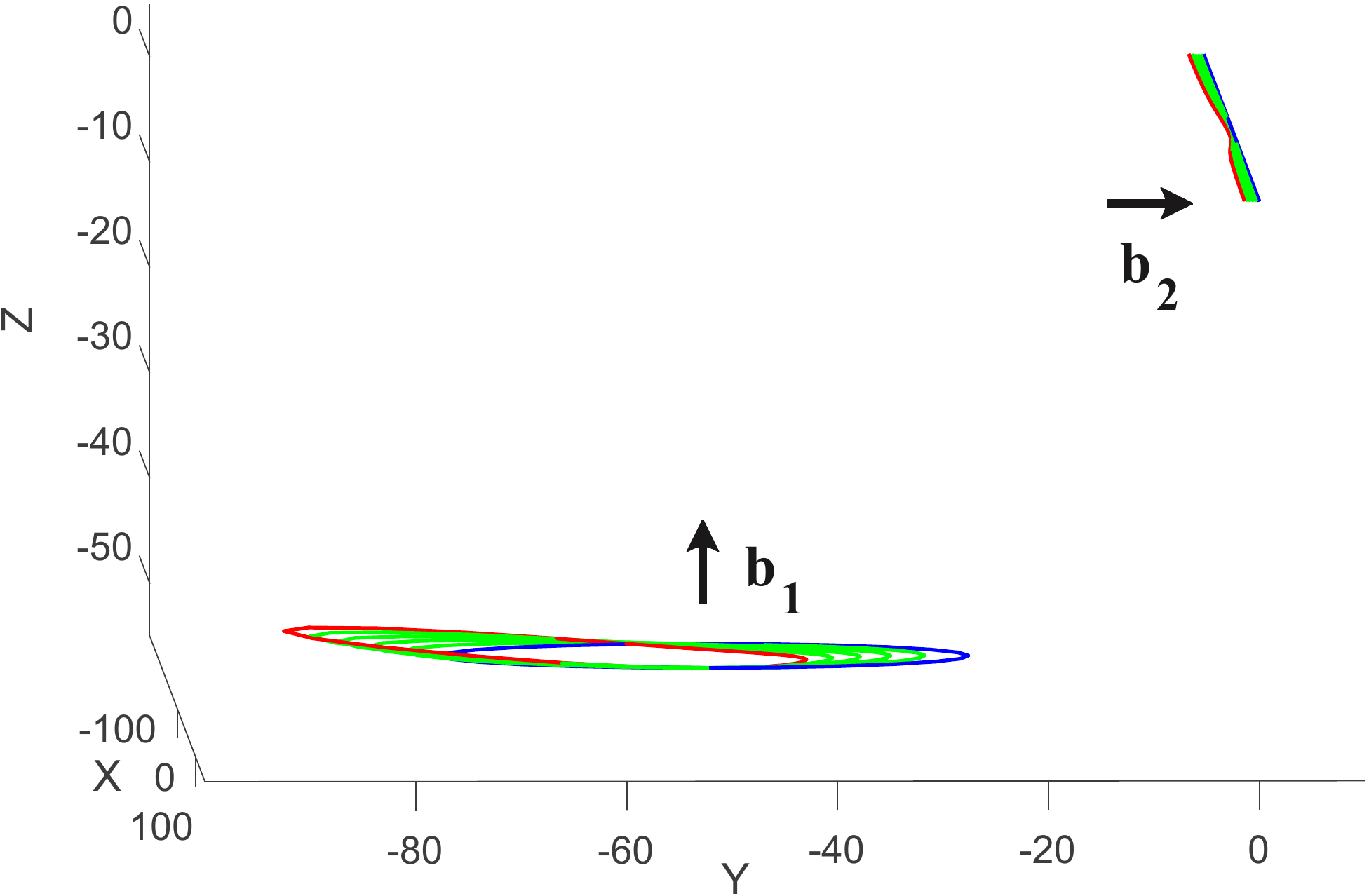}}  \hspace{1in}
\subfigure[]{\includegraphics[width=1in]{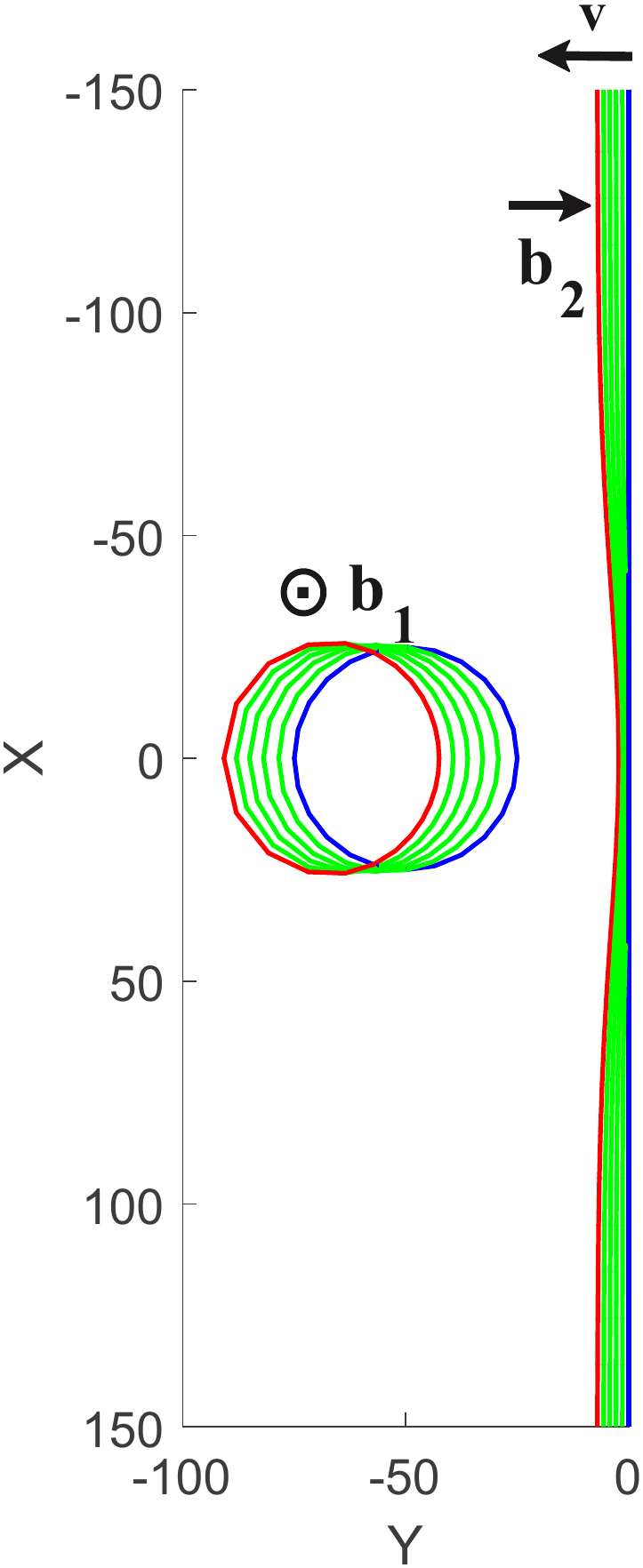}}  \hspace{1in}
\subfigure[]{\includegraphics[width=2.4in]{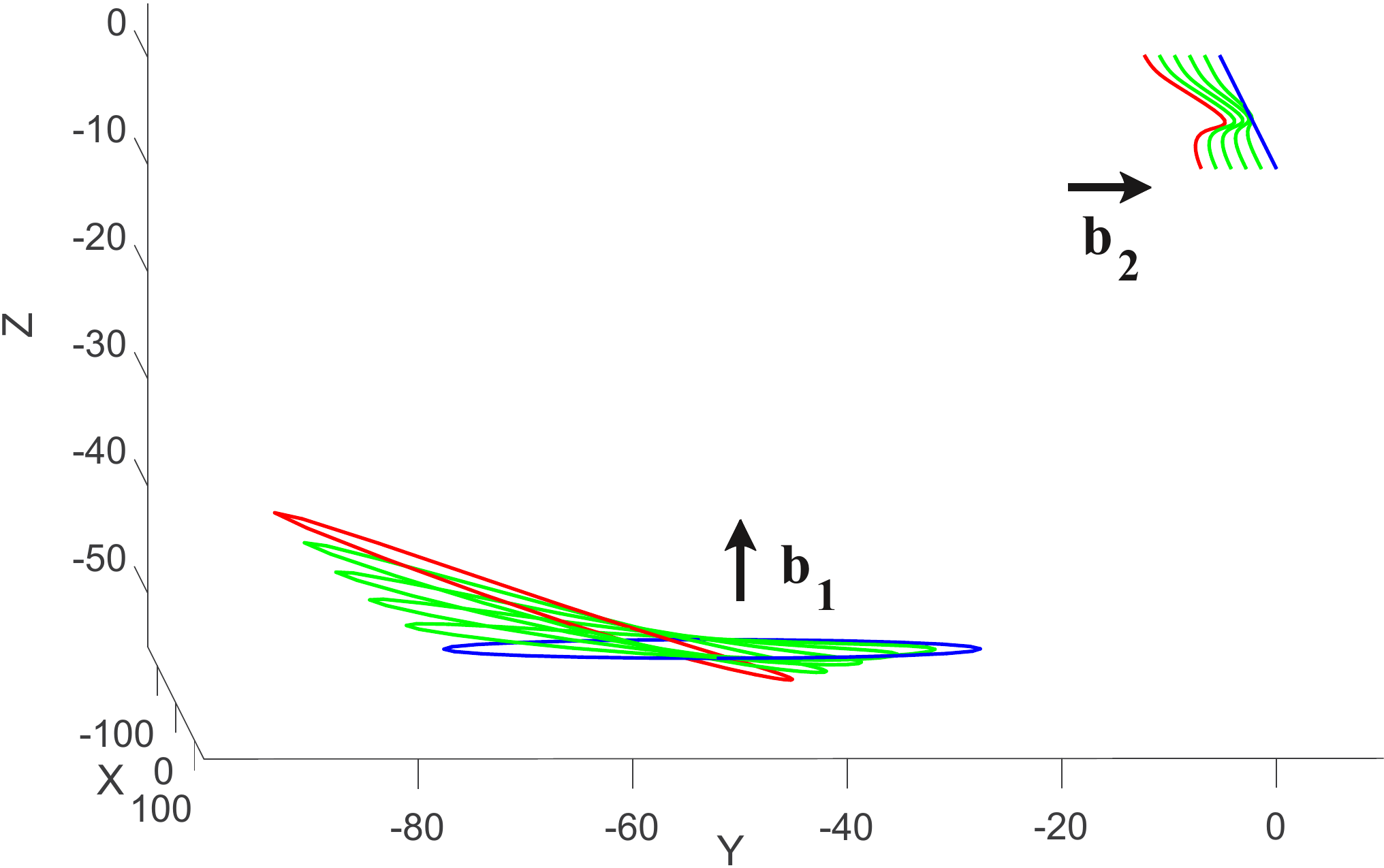}}\hspace{1in}
\subfigure[]{\includegraphics[width=1in]{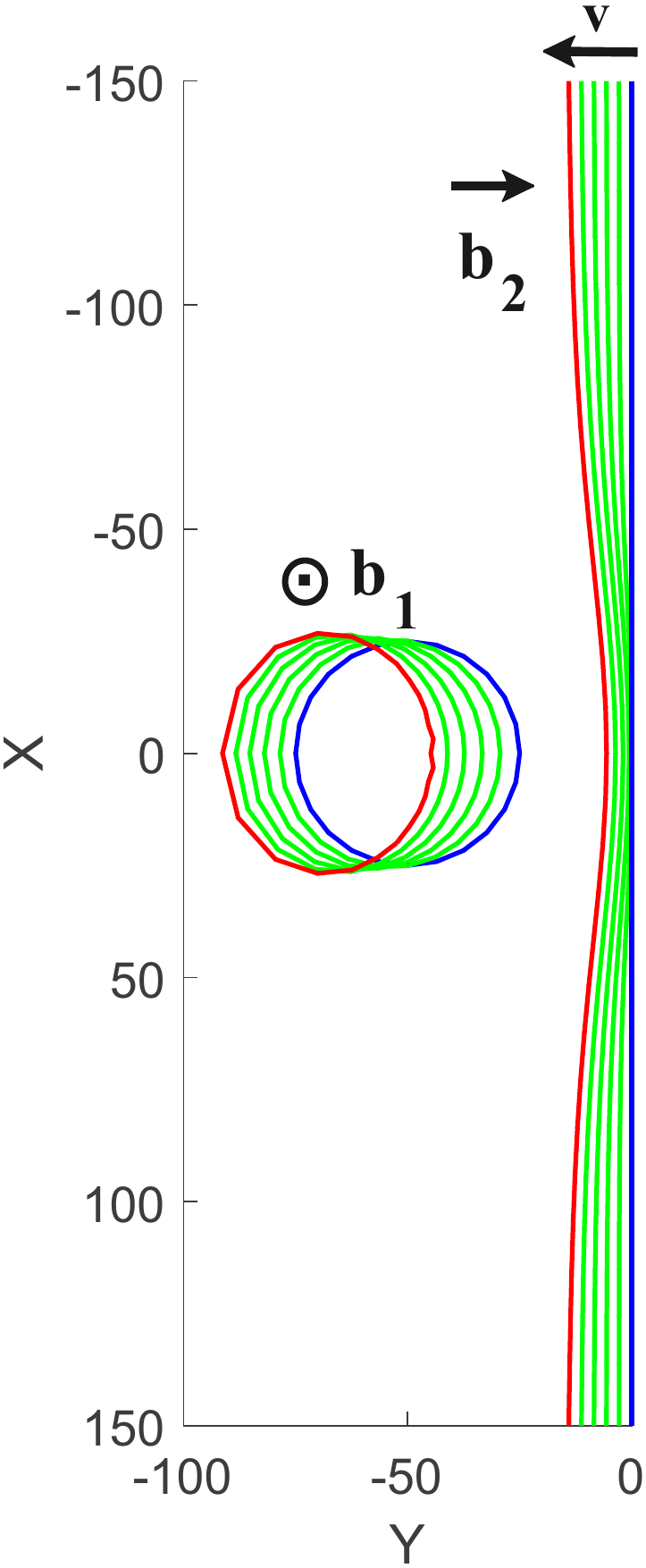}}\hspace{1in}
\subfigure[]{\includegraphics[width=2.4in]{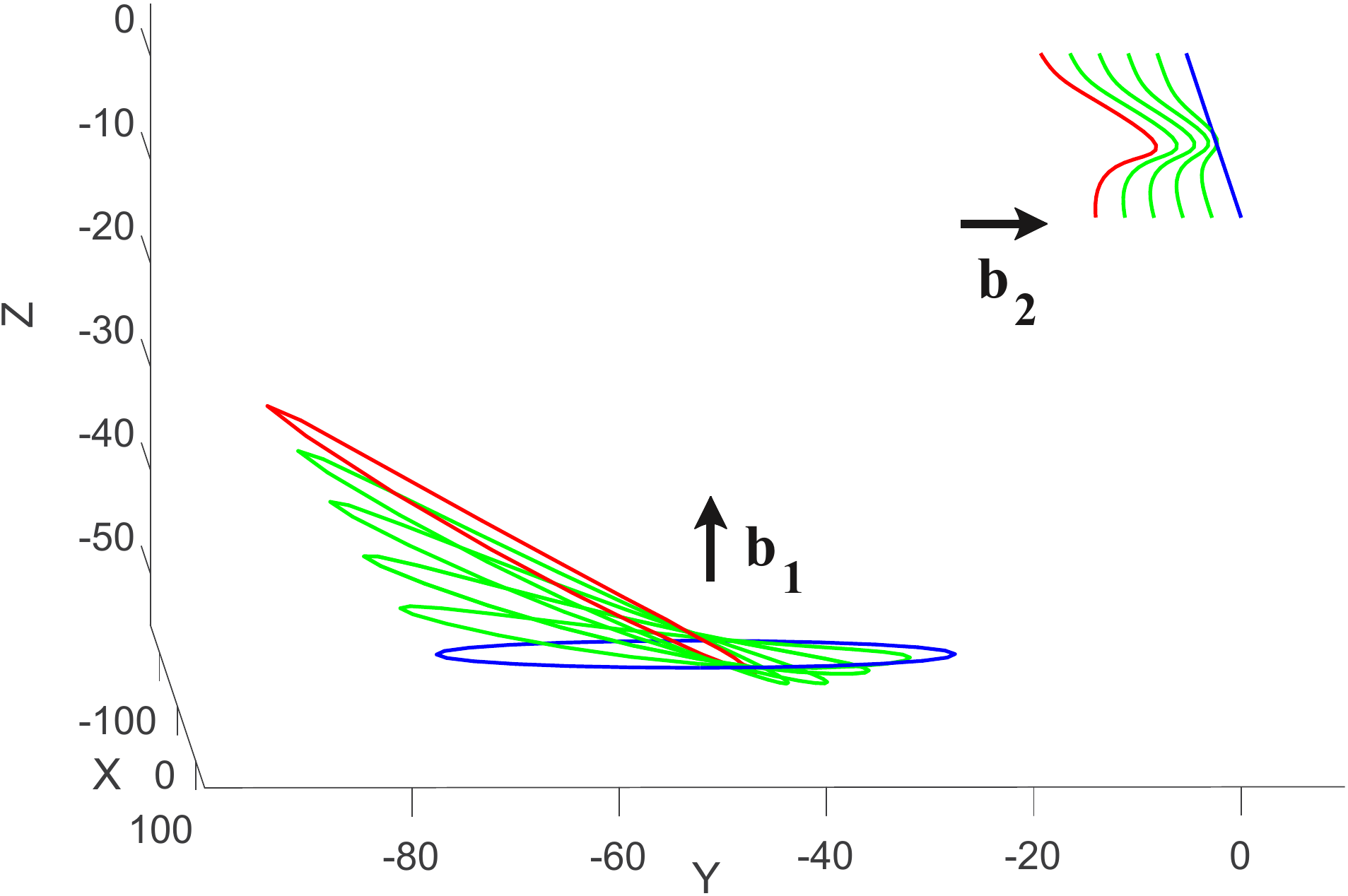}}
\caption{Simulations of translation of a prismatic loop by self climb driven by the stress field of a moving edge dislocation. The edge dislocation moves by glide with an applied stress $\sigma_{23}^{\text{app}}=-10^{-3}\mu$.  The glide mobility $M_{\rm g}/(c_0D_c/\mu b)=0.1$, $0.5$, and $1$ in the top, middle, and bottom panels, respectively. The length unit is $b$ in these images.
(a), (c), and (e): Snapshots of the prismatic loop and edge dislocation during the evolution viewed from the top ($z$ direction). (b), (d), and (f): Three dimensional view of the snapshots during the evolution. In each simulation (shown in each panel), configuration with blue ad red colors are the initial state and the last state of the simulation, respectively.   The elapsed time  is $t= 8000 b^2/(c_0D_c)$, and the snapshots are equally spaced in time. }\label{fig.loopandedge1}
\end{figure}

 In this subsection, we present simulations for the translation of a prismatic dislocation loop  by self-climb driven by the stress field of a moving  edge dislocation.
The initial loop is a circular prismatic loop with radius $R=25b$ and center $(0,-50b,-50b)$ and is parallel to  the $xy$ plane. The Burgers vector of this loop is $b_1=(0, 0, b)$.
 An edge dislocation is located on the  $x$ axis and is in $+x$ direction.  The Burgers vector of this edge dislocation is $b_2=(0, b, 0)$.  See Fig.~\ref{fig.loopandedge1} for the initial configuration.
 In the simulation, the edge dislocation is tracked within $[-150b,150b]$ in the $y$ direction and interactions with straight extension of $150b$ from each side are kept.
 Under an applied stress $\sigma_{23}^{\text{app}}=-10^{-3}\mu$, the edge dislocation moves in the $-y$ direction, towards the prismatic loop, by glide motion.
 The glide velocity is calculated by $v_{\text{g}}=M_{\rm g}f_{\text{g}}$, where $f_{\text{g}}$ is the glide component of the Peach-Koehler force $\mathbf{f}_{\rm PK}$ (that in the slip plane). In the simulations, we choose three values of the glide mobility $M_{\rm g}/(c_0D_c/\mu b)=0.1$, $0.5$, and $1$.
  The simulation configuration is to mimic that of the experiment in Zinc conducted by  Kroupa and Prince in 1961 \cite{Kroupa1961}.

The DDD simulation results using our formulation in Eq.~\eqref{eq:climb-v} are shown in Fig.~\ref{fig.loopandedge1}. In each simulation, as the edge dislocation glides towards the prismatic loop under the applied stress, the prismatic loop is pushed by self-climb due to the stress field generated by the edge dislocation. These simulations agree excellently with the experimental observation by  Kroupa and Prince \cite{Kroupa1961}. Note that during the evolution, the self-climb motion of the prismatic loop is not a rigid translation; its shape changes as it evolves. As the prismatic loop translates, it also rotates by glide due to the applied stress and the stress field generated by the edge dislocation.
 The edge dislocation becomes  curved due to the interaction with the prismatic loop. For relatively small glide mobility  $M_{\rm g}/(c_0D_c/\mu b)=0.1$ (or relatively fast pipe diffusion), the rotation of the prismatic loop is not that significant as it translates by self-climb; for glide mobility  $M_{\rm g}/(c_0D_c/\mu b)=0.5$, the prismatic loop rotates more as it translates; and for the relatively large glide mobility  $M_{\rm g}/(c_0D_c/\mu b)=1$ (or relatively slow pipe diffusion), the rotation of the prismatic loop is significant as it translates. The edge dislocation moves more as the glide mobility increases within the same time period.  As will be shown  in the next section, there is no stable location of the prismatic loop under motions of self-climb and glide.   Fig.~\ref{fig.loopandedge1} shows simulation results up to time $t= 8000 b^2/(c_0D_c)$.

\section{Interaction between a prismatic loop and an edge dislocation}\label{sec:interaction}

In this section, we perform analyses to systematically understand the behaviors of a prismatic loop under the elastic interaction with an infinite, straight edge dislocation by the motion of self-climb and glide. These analyses also help to understand the translation of a prismatic loop by self-climb driven by a moving edge dislocation presented in Sec.~\ref{subsec:movingedge}.

\begin{figure}[htbp]\label{fig.EdgeLoopMoveDirection}
\centering
\subfigure[]{\label{fig.edgeloop1}\includegraphics[width=2.5in]{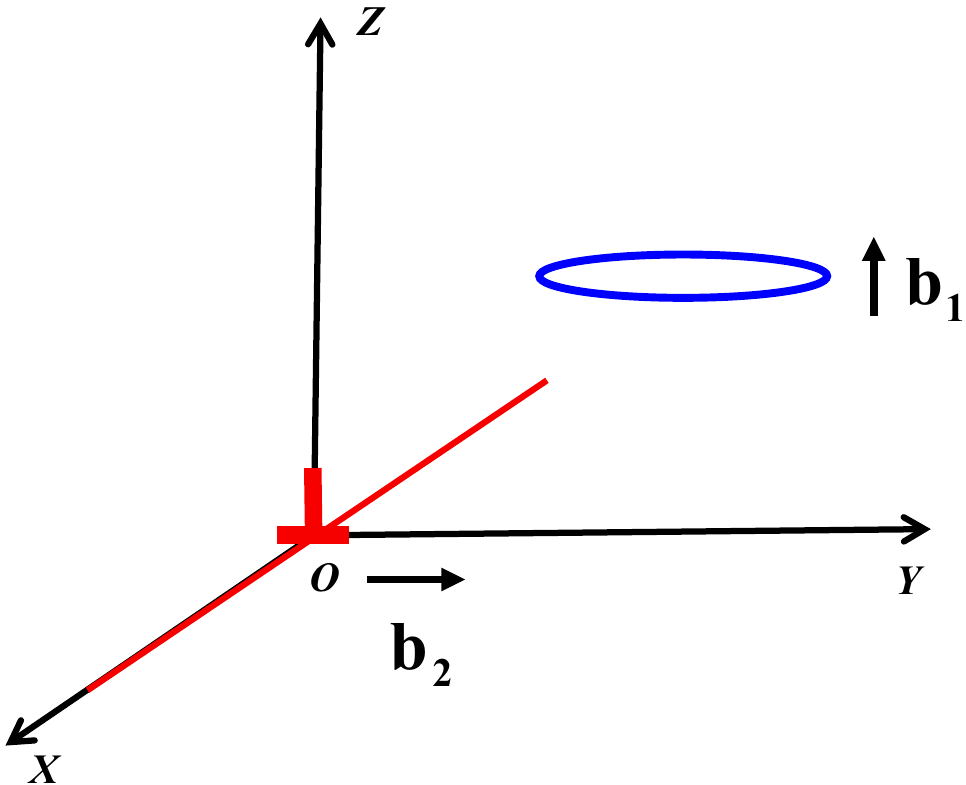} } \vspace{0.5in}\\
\subfigure[]{\label{fig.edgeloop3} \includegraphics[width=5.5in]{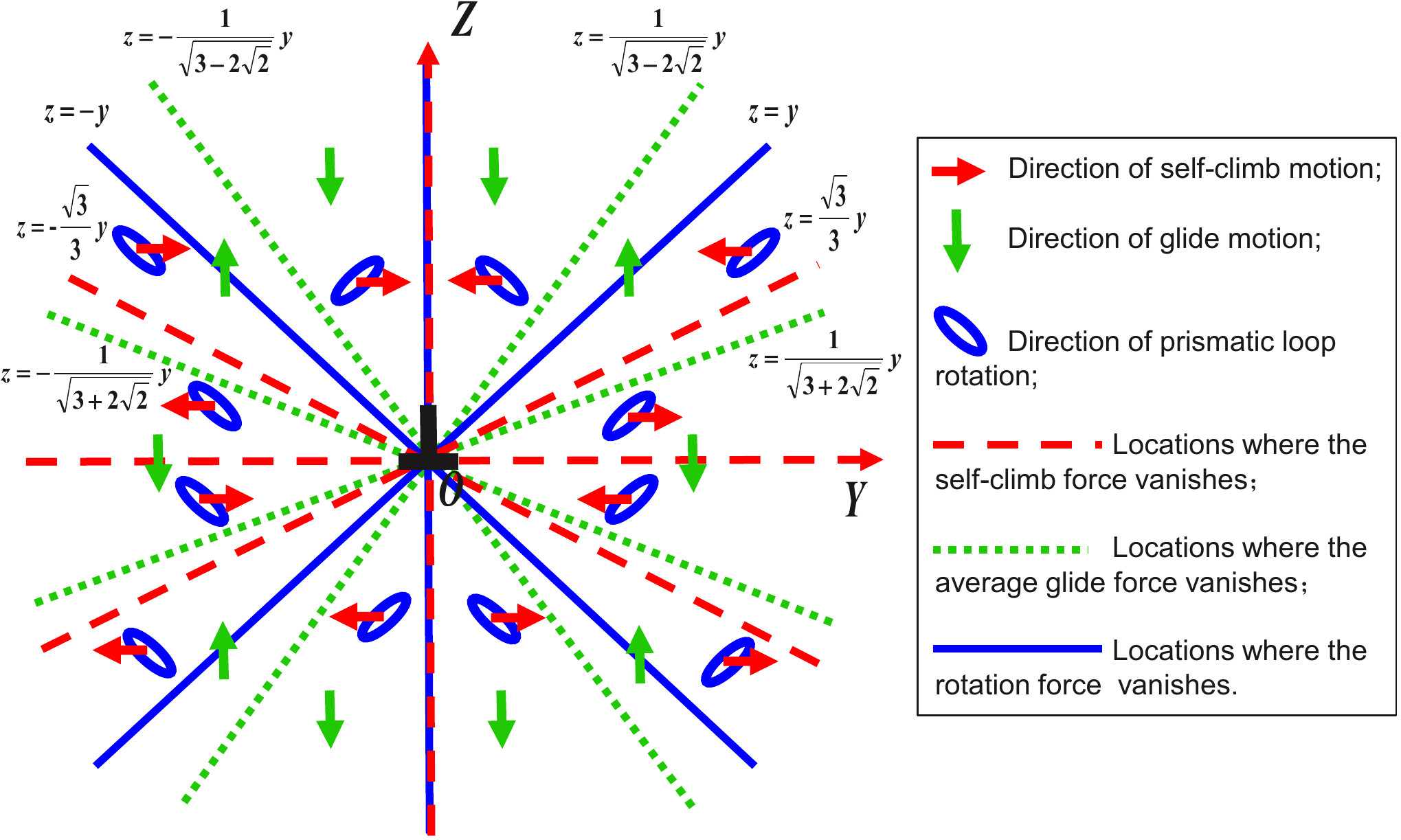}} \vspace{0.5in}
\caption{(a) Configuration of a small prismatic dislocation loop and an infinite edge dislocation. (b)
Directions of the motions of self-climb, glide, and rotation of the prismatic loop under the stress field of the edge dislocation, illustrated by red arrows, green arrows, and rotated blue loops, respectively. The red, green, and blue lines show locations where the self-climb force, the average glide force, and the rotation force on the prismatic loop vanish, respectively.
  }
\end{figure}

We assume that an infinite straight edge dislocation is located on the $x$ axis with Burgers vector $(0,b_2, 0)$ and is in the $+x$ direction.   We examine the behaviors of a  small circular prismatic dislocation loop with Burgers vector $(0,0,b_1)$ under the stress field of the edge dislocation. The prismatic loop is in the counterclockwise direction and its radius  is $R$. See Fig.~\ref{fig.edgeloop1} for the configuration.

From Eq.~\eqref{eq-average-velocity0}, the translation velocity of the small prismatic loop with center located at $(x,y,z)$ can be approximately given by
\begin{eqnarray}\label{eq-average-velocity1}
\bar{\mathbf{v}}_{\rm cl}=-\frac{c_0 D_c e^{\frac{\sigma_{33}^0\Omega}{k_BT}}\Omega b_1}{k_BTR}
\left(\frac{\partial \sigma_{33}}{\partial x}, \frac{\partial \sigma_{33}}{\partial y},0\right)
=\frac{c_0 D_ce^{\frac{\sigma_{33}^0\Omega}{k_BT}}\Omega b_1}{k_BTR}\frac{\mu b_2}{2\pi(1-\nu)}\left(0,\frac{2yz(y^2-3z^2)}{(y^2+z^2)^3},0\right),
\end{eqnarray}
where $\pmb\sigma$ is the stress generated by the edge dislocation, $\sigma_{33}=\frac{\mu b_2}{2\pi(1-\nu)}\frac{z(y^2-z^2)}{(y^2+z^2)^2}$, and the value of its gradient is evaluated at the center of the prismatic loop.
The glide force on the prismatic loop due to the stress field generated by the edge dislocation is
\begin{eqnarray}\label{eq.normal.glide}
f_{\text{g}}=-\xi_1b_1\sigma_{23}=-\frac{\xi_1\mu b_1b_2}{2\pi(1-\nu)}\frac{y(y^2-z^2)}{(y^2+z^2)^2},
\end{eqnarray}
where the unit tangent vector at a point on the prismatic loop is $\boldsymbol{\xi}=(\xi_1,\xi_2,0)$.
The average glide force on the small prismatic loop is approximately
\begin{eqnarray}\label{eq.normal.glide.loop}
\bar{f}_{\text{g}}^{\text{loop}}=\frac{1}{2\pi R}\int_\gamma f_{\text{g}}ds =-\frac{\mu b_1b_2R}{8\pi(1-\nu)}\cdot\frac{y^4-6y^2z^2+z^4}{(y^2+z^2)^3},
\end{eqnarray}
where $\gamma$ is the prismatic loop. The glide velocities of the loop are calculated by $v_{\rm g}=M_{\rm g}f_{\rm g}$ and $\bar{v}_{\rm g}^{\rm loop}=M_{\rm g}\bar{f}_{\rm g}^{\rm loop}$, where $M_{\rm g}$ is the glide mobility.

We analyze the behaviors of the small prismatic loop by motions of self-climb, glide, and rotation
under the stress field of the edge dislocation. The directions of different kinds of motions of the prismatic loop  as well as the locations where these forces (velocities) vanish are shown in Fig.~\ref{fig.edgeloop3}. These results are based on Eqs.~\eqref{eq-average-velocity1}--\eqref{eq.normal.glide.loop}. In particular, the rotation tendency of the loop is determined by Eq.~\eqref{eq.normal.glide} at the two ends of the loop that are closest and farthest to the edge dislocation, where $\xi_1=-1$ or $1$.

We focus on possible stable locations of the prismatic loop under various forces. The self-climb velocity (force) on the loop vanishes when the center of the loop is located on  $z=\pm\frac{3}{\sqrt{3}}y$ and $y=0$. The self-climb equilibrium location is stable when $z=-\frac{3}{\sqrt{3}}y$ with $y>0$, $z=\frac{3}{\sqrt{3}}y$ with $y<0$, and $y=0$ with $z>0$. When the self-climb motion of the loop is faster than the glide motion, the loop may first reach a stable location with respect to self-climb and then further moves by glide.
On a self-climb stable location of $z=-\frac{3}{\sqrt{3}}y$ with $y>0$, the glide force is upward. As a result, the loop will go upward by glide first and then is attracted to the $z=-\frac{3}{\sqrt{3}}y$ line to the left by self-climb, leading to a zigzag motion towards the edge dislocation.
The behavior of the loop is similar when it is located on the self-climb stable location of $z=\frac{3}{\sqrt{3}}y$ with $y<0$. Finally, on a self-climb stable location of $y=0$ with $z>0$, the glide force on the loop is downward, and the loop will go downward by glide along this self-climb stable line. Note that from Eqs.~\eqref{eq.normal.glide} and \eqref{eq.normal.glide.loop}, the average glide force on the prismatic loop $\bar{f}_{\text{g}}^{\text{loop}}$ is of order $O(R/L)$ of the glide force $f_{\text{g}}$ on a point on the loop when the loop radius $R$ is small compared with its distance $L$ to the edge dislocation. The above behaviors of the loop still hold when the glide velocity $v_{\text{g}}$ is comparable or not very larger than the self-climb velocity $\bar{v}_{\rm cl}$. As discussed above, there is no stable location of the prismatic loop under both motions of self-climb and glide in this regime of fast self-climb.

When the glide motion of the prismatic loop is much faster than its self-climb motion, the loop may first reach a stable location with respect to glide and then further moves by self-climb.
The glide velocity (force) on the loop vanishes when the center of the loop is located on  $z=\pm\frac{1}{\sqrt{3\pm 2\sqrt{2}}}y$.
The glide equilibrium location is stable on the following four rays:
$z=\frac{1}{\sqrt{3-2\sqrt{2}}}y$ with $y>0$,
$z=-\frac{1}{\sqrt{3+ 2\sqrt{2}}}y$ with $y>0$,
$z=-\frac{1}{\sqrt{3- 2\sqrt{2}}}y$ with $y<0$, and
$z=\frac{1}{\sqrt{3+2\sqrt{2}}}y$ with $y<0$. In all glide stable locations, the self-climb velocity is nonzero and is towards the $z$ axis. This make the loop go toward the edge dislocation by a zigzag motion along these glide stable rays under both the glide and self-climb motions, similar to the discussion above in the fast self-climb case. Again, there is no stable location of the prismatic loop under both motions of self-climb and glide in this regime of fast glide.

The prismatic loop can also rotate by glide along its slip cylinder under the stress field of the edge dislocation and the rotation will be hindered by its self-force~\cite{Kroupa1966,Wolfer2004}. Fig.~\ref{fig.edgeloop3} also shows the rotation direction of the small prismatic loop. The glide force for rotation vanishes when $z=\pm y$ and $y=0$.

Now we understand the translation of a prismatic loop by self-climb driven by a moving edge dislocation  presented in Sec.~\ref{subsec:movingedge}. Initially, the center of the prismatic loop is located at $y=z=-50b$. As can be see from Fig.~\ref{fig.edgeloop3}, at this location, the loop is pushed away by the stress field of the edge dislocation by self-climb motion; this also makes the loop enter the region $y<z<0$, where the rotation of the loop is in such a way that its far end goes up and the near end goes down. The average glide velocity of the loop is upward as shown in Fig.~\ref{fig.edgeloop3}. These properties perfectly explain the simulation results of the prismatic loop as an edge dislocation is driven towards the loop presented in Sec.~\ref{subsec:movingedge}.

Finally, we would like to remark that when line direction of the prismatic loop changes from the current counterclockwise direction to clockwise (or Burger vector of the loop/Burgers vector of the edge dislocation/line direction of the edge dislocation changes to its opposite direction), the new figure of directions of different kinds of motions will be mirror reflection of Fig.~\ref{fig.edgeloop3} with respect to the $y$ axis.

%

\section{Conclusions and Discussion}

In this paper, we have shown that our self-climb DDD formulation in Eq.~\eqref{eq:climb-v} (or \eqref{eq:climb-v0}) is able to quantitatively describe the properties of self-climb of prismatic loops that were observed in experiments and atomistic simulations. This dislocation dynamics formulation applies to self-climb by pipe diffusion for any configurations of dislocations.
In particular,  with our formulation, we show the following properties of self-climb of prismatic loops. The area enclosed by a prismatic loop is conserved during the self-climb motion; the equilibrium shape of a prismatic loop with the self-climb motion under its self stress is a circle; and the circular shape is preserved in the self-climb motion under a constant stress gradient field.
Using our DDD self-climb model, we have also obtained a velocity formula for translation of a circular prismatic loop by self-climb under constant stress gradient. Under the conditions that the stress gradient is small or the radius of the prismatic loop is small, our translation velocity formula recovers the available theories in the literature for the self-climb models based on rigid translation of a circular prismatic loop \cite{Johnson1960,Turnbull1970,Narayan1972,Swinburne2016,Okita2016}. Our formulation provides a DDD justification of these rigid translation models in the above regimes; otherwise in a general case,  our pointwise DDD self-climb formulation should be used.
The parameters in our self-climb DDD formulation can be estimated by comparisons with experimental results \cite{Johnson1960,Turnbull1970,Narayan1972,Swinburne2016} or atomistic simulations \cite{Swinburne2016,Okita2016}.

We have also presented a DDD implementation method for this self-climb formulation. Simulations performed show evolution, translation  and coalescence of prismatic loops as well as the translation of prismatic loops by self-climb driven by a moving edge dislocation. These results are in excellent agreement with available experimental observations on translation of prismatic loops by self-climb driven by a moving edge dislocation~\cite{Kroupa1961} and coalescence of prismatic loops \cite{Silcox1960,Washburn1960,Johnson1960,Turnbull1970,Narayan1972,Swinburne2016}.
Our DDD simulations have also validated the properties and analytical formulas of the self-climb motion obtained in this paper and summarized above. We have also performed analyses to systematically understand the behaviors of a prismatic loop under the elastic interaction with an infinite, straight edge dislocation by motions of self-climb and glide (including both rotation and rigid motions by glide) of the prismatic loop.

 Our dislocation self-climb formulation  can be easily incorporated in the available DDD simulation methods and codes for the  motions of glide and climb assisted by vacancy bulk diffusion, as an additional contribution to the dislocation velocity, which enables simulations on the combined effects of dislocation motions of glide, climb by vacancy bulk diffusion, and self-climb by vacancy pipe diffusion. Future work may also include investigations on the coupling effects with irradiation \cite{Was,DudarevReview}, vacancy diffusion influenced by elastic effect \cite{Hirth-Lothe}, and influence of moving dislocations \cite{Krasnikov}.

\section*{Acknowledgments}
This work was partially supported by the Hong Kong Research Grants Council General Research Fund 16302115, and National Natural Science Foundation of China under the grant number  11801214.

\bibliographystyle{plain}

\bibliography{mybib}

\end{document}